\documentclass[a4paper,11pt]{article}
\pdfoutput=1 % if your are submitting a pdflatex (i.e. if you have
\usepackage{jinstpub} % for details on the use of the package, please
\usepackage{color}
\usepackage{xcolor}
\usepackage{float}
\usepackage{xspace}
\usepackage{upgreek}
\usepackage{subcaption}
\usepackage{lineno}
\newcommand{\xfel}{European~XFEL\xspace}
\newcommand{\pandora}{PANDORA\xspace}

\newcommand{\csitl}{CsI(Tl)\xspace}
\newcommand{\purecsi}{pure CsI\xspace}
\newcommand{\Purecsi}{Pure CsI\xspace}

\title{Energetic neutron identification with pulse shape discrimination in \purecsi crystals}

\author[a]{S. Longo}
\affiliation[a]{Deutsches Elektronen-Synchrotron DESY, Notkestr. 85, 22607 Hamburg, Germany}

\author[a, b, c]{, M. Khan}
\affiliation[b]{Institut f\"ur Experimentelle Teilchenphysik,  Karlsruher Institut f\"ur Technologie (KIT), 76131 Karlsruhe, Germany}

\author[c, d]{, S. Dharani}
\affiliation[c]{Institut f\"ur Experimentalphysik, Universit\"at Hamburg, 22761 Hamburg, Germany}

\author[d]{, B. von Krosigk}
\affiliation[d]{Institut f\"ur Astroteilchenphysik, Karlsruher Institut f\"ur Technologie (KIT), 76344 Eggenstein-Leopoldshafen, Germany}

\author[b]{and T. Ferber}

\emailAdd{savino.longo@desy.de}

\abstract{Pulse shape discrimination with \purecsi scintillators is investigated as a method for separating energy deposits by energetic neutrons and photons at particle physics experiments. Using neutron data collected near the \xfel XS1 beam window the pulse shape discrimination capabilities of \purecsi are studied and compared to \csitl using near-identical detector setups, which were operated in parallel. The inelastic interactions of 100\,MeV neutrons are observed to produce a slower scintillation emission in \purecsi relative to energy deposits from cosmic muons.  By employing a charge-ratio method for pulse shape characterization, pulse shape discrimination with \purecsi is shown to be effective for identifying energy deposits from neutrons vs. cosmic muons, however, \purecsi was not able resolve the specific type of neutron inelastic interactions as can be done with \csitl.      Using pulse shape discrimination, the rate of energetic neutron interactions in a \purecsi detector is measured as a function of time and shown to be correlated with the \xfel beam power.  The results demonstrate that pulse shape discrimination with \purecsi has significant potential to improve electromagnetic vs. hadronic shower identification at future particle physics experiments.}


\begin{document}
\maketitle
\flushbottom

\section{Introduction}

Crystal calorimeters are frequently employed by particle physics experiments to perform photon detection and energy measurement, neutral hadron detection, and charged-particle identification.  Recent innovations in crystal calorimetry have expanded the information recorded from each calorimeter crystal to extend beyond the crystal energy and timing by also including the pulse shape of the scintillation emission \cite{Sugiyama:2021ltp,Longo:2020zqt}. The new pulse shape information can be used to identify whether an energy deposit is from an electromagnetic or hadronic shower.  This is achieved by exploiting the particle-dependent scintillation response present in many inorganic scintillators \cite{Longo:2018uyj}. Using pulse shape discrimination with CsI(Tl), the Belle II experiment \cite{abe2010belle,Belle-II:2018jsg} has demonstrated significant improvements to the calorimeter's capabilities for photon vs. $K^0_L$ separation \cite{Longo:2020zqt}. Pulse shape discrimination with CsI(Tl) is also applied in recent charged particle detector systems such as FAZIA \cite{Barlini_2020} and ChAKRA \cite{KUNDU2019162411}.
 Additionally, the KOTO experiment, which employs a pure CsI crystal calorimeter, has demonstrated that pulse shape discrimination with pure CsI allows for improvements in photon vs. neutron \mbox{separation \cite{Sugiyama:2021ltp}}.  

Pulse shape discrimination has been well-established in several inorganic scintillators, such as CsI(Tl) and NaI(Tl) \cite{Storey1958,BARTLE199954,Longo:2018uyj,DINCA2002141}, there remains however many frequently used inorganic scintillators where the pulse shape discrimination capabilities are less established.  Illustrating this, the recent report by the KOTO experiment \cite{Sugiyama:2021ltp} was the first demonstration of energetic neutron vs. photon separation using pulse shape discrimination in pure CsI.  The fast timing and radiation hardness of pure CsI \cite{pureCsIRadHard2016,KUBOTA1988275,Aihara_FastCalor} allow it to find application at several recent particle and nuclear physics experiments including KOTO \cite{Sugiyama:2021ltp}, PIENU \cite{PIENU_PhysRevLett.115.071801}, and PIBETA \cite{PIBETA_2004} as well as planned use in the future experiments Mu2e \cite{mu2e_2018} and the Super Charm-Tau Factory \cite{Barniakov:2019zhx,EIDELMAN2015238,Aihara_FastCalor}.   In this paper the pulse shape discrimination capabilities of pure CsI are explored further by comparing the scintillation response to energy deposits from energetic neutrons and cosmic muons. Using a charge-ratio method for pulse shape characterization, the ability for pulse shape discrimination in pure CsI to isolate energy deposits from energetic neutrons is demonstrated.

This paper is organized as follows.  Section \ref{sec2_ExpSetup} details the experimental setup and outlines the data samples collected at the European X-Ray Free-Electron Laser (\xfel) Facility. In Section \ref{sec_PSD} the pulse shape discrimination capabilities of pure CsI are studied and compared to CsI(Tl).  Section \ref{sec_monitoring} demonstrates that using pulse shape discrimination, the rate of energetic neutrons produced near the \xfel beam dump can be isolated from a cosmic muon background and measured with a performance similar to commercially available energetic neutron detection systems.  Conclusions are presented in Section \ref{sec_conclusion} summarizing the results and future directions and applications for pulse shape discrimination in pure CsI.

\section{Experimental Details}
\label{sec2_ExpSetup}

\subsection{\Purecsi and \csitl Detectors}

\begin{figure}[ht]
  \centering
\includegraphics[width=0.65\textwidth]{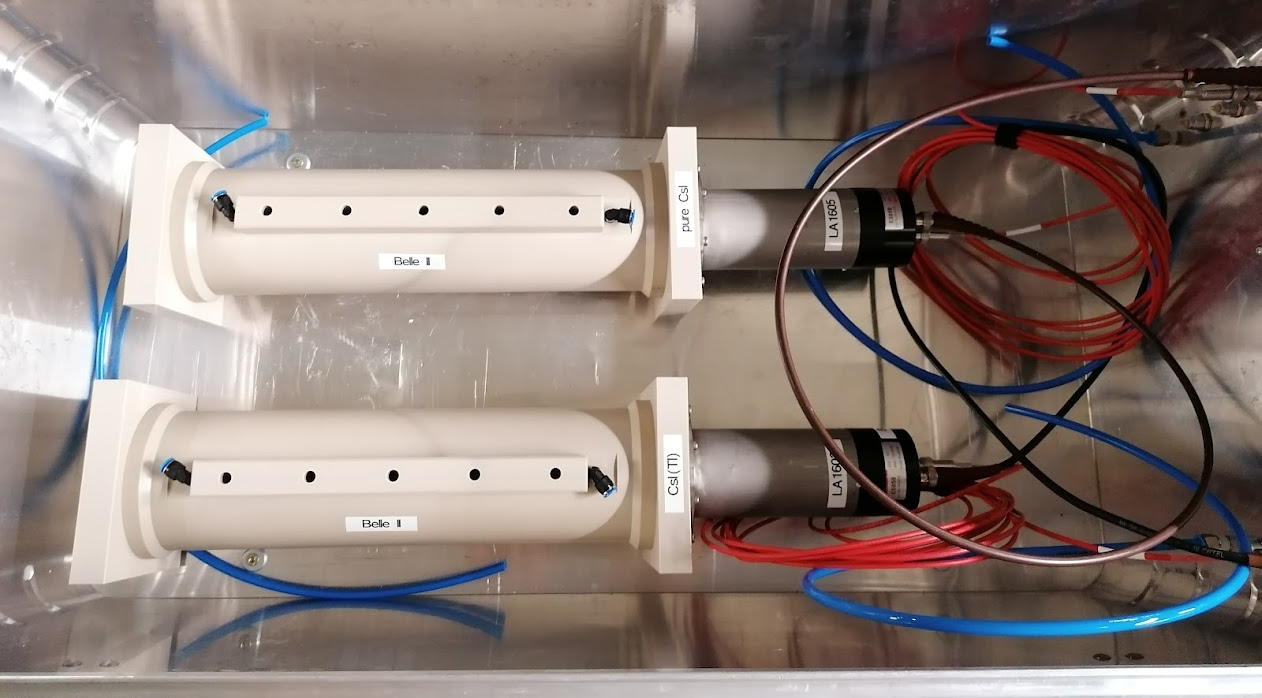}\\
\caption{Photo of the \purecsi and \csitl detector assemblies. }
\label{fig_ExperimentalSetup}
\end{figure}

The experimental setup is shown in Figure \ref{fig_ExperimentalSetup} and consists of two identical detector assemblies, one containing a \purecsi crystal and the other a \csitl crystal.  
The \purecsi and \csitl crystals were manufactured by St.~Gobain and have a rectangular prism geometry with dimensions $5 \times 5 \times 30$\,cm$^3$, similar to the large crystal sizes used in particle physics calorimeters \cite{abe2010belle,Sugiyama:2021ltp,mu2e_2018}.
Both detectors use a Hamamatsu R5113-02 photomultiplier tube (PMT) mounted at one end of the crystal for scintillation light detection. This PMT is equipped with a UV-transparent window matched to the near UV emission of \purecsi \cite{PMTdatasheet}. The optical connection between the crystal and PMT corresponds to an air gap. Both crystals have a thin Teflon wrapping to improve light collection. The analog output of the PMT is digitized over a 30\,$\upmu$s time window with 14\,bit precision by a CAEN DT5730SB digitizer with sampling time of 2\,ns~\cite{DigitzerDatasheet}. 
Both detector assemblies are enclosed in a light tight housing and are operated simultaneously.
The digitizer is operated in self-triggering mode on a constant voltage threshold, triggering independently for the \purecsi and \csitl crystal. If the signal from one of the crystals activates the trigger, then only its digitized waveform was recorded for offline analysis. 

The commercially available LB6419 neutron and gamma dose rate monitoring system, referred to as the \pandora detector and used by the \xfel for radiation monitoring \cite{KLETT20101242}, was also operated in parallel and located directly adjacent to the \purecsi and \csitl detector systems.  High energy neutron detection is achieved with \pandora using a plastic scintillator with cylindrical geometry that has a diameter of 48\,mm and length of 48\,mm \cite{KLETT20101242,PandoraDatasheet,PANDORApaper}. 

\subsection{Experimental data}
\label{sec_Experimentaldata}

The experimental setup is positioned in the UG2 area of the \xfel, located approximately 10.3\,m above XS1 beam dump window. Between the detector setup and the beam window is a large air gap followed by 30\,cm of heavy concrete (Density = 3.8\,g/cm$^3$) followed by another large air gap and then 2\,m of concrete (Density = 2.3\,g/cm$^3$)~\cite{DESYInternalReport}.  Photon backgrounds originating from the beam window region are suppressed by the concrete shielding resulting in a negligible energetic photon background in the region of the detectors \cite{DESYInternalReport}.  During \xfel beam operation, the 17.5\,GeV electrons passing through the dump window generate a flux of energetic neutrons in the region of the experimental setup.  The neutrons have an energy spectrum that is peaking approximately at a kinetic energy of 100\,MeV \cite{DESYInternalReport}.  Additional information on the neutron and photon flux present the UG2 area during XFEL operation can be found in reference \cite{DESYInternalReport}. In addition to the energetic neutrons, a continuous flux of cosmic muons is also always present.  Over the period of several weeks, two datasets were accumulated and referred throughout the remainder of the text as the \textit{beam-off} dataset and \textit{beam-on} dataset. The beam-off dataset corresponds to digitized waveforms recorded while the \xfel beam was not in operation and as a result corresponds to energy deposits primarily from cosmic muons ionizing through the crystal.   The beam-on dataset was accumulated while the XFEL beam was in operation, and contains waveforms from energy deposits by both cosmic muons and energetic neutrons.

\subsection{Waveform pre-selection}

A pre-selection is applied offline to remove waveforms that exceeded the maximum ADC voltage range.  Waveforms with pile-up, corresponding to multiple scintillation peaks in the same time window are also removed by applying the following algorithm. Using the \texttt{find\_peaks} method from the \texttt{scipy} python package \cite{scipy,2020SciPy-NMeth} the number of local minima present in the waveform's 30\,$\upmu$s time window is measured.  The \texttt{find\_peaks} prominence parameter is set to 8.5\% for \purecsi and 40\% for \csitl.  Waveforms with two local minima within the 30\,$\upmu$s time window that are separated by a timing window of $\geq 30$\,ns for \purecsi and $\geq 0.4$\,$\upmu$s for \csitl are classified as pile-up waveforms and rejected.  Over 98\% of pure CsI and CsI(Tl) waveforms in the beam-off sample pass the pre-selection.  In the beam-on sample, over 90\% of pure CsI waveforms and 96\% of CsI(Tl) waveforms pass the pre-selection.  The waveform energy distributions before and after the pre-selection are included in Appendix A.  Waveforms are also required to have a total energy above 12\,MeV.

\begin{figure}[ht]
\begin{subfigure}{1\textwidth}
  \centering
\includegraphics[width=0.49\textwidth]{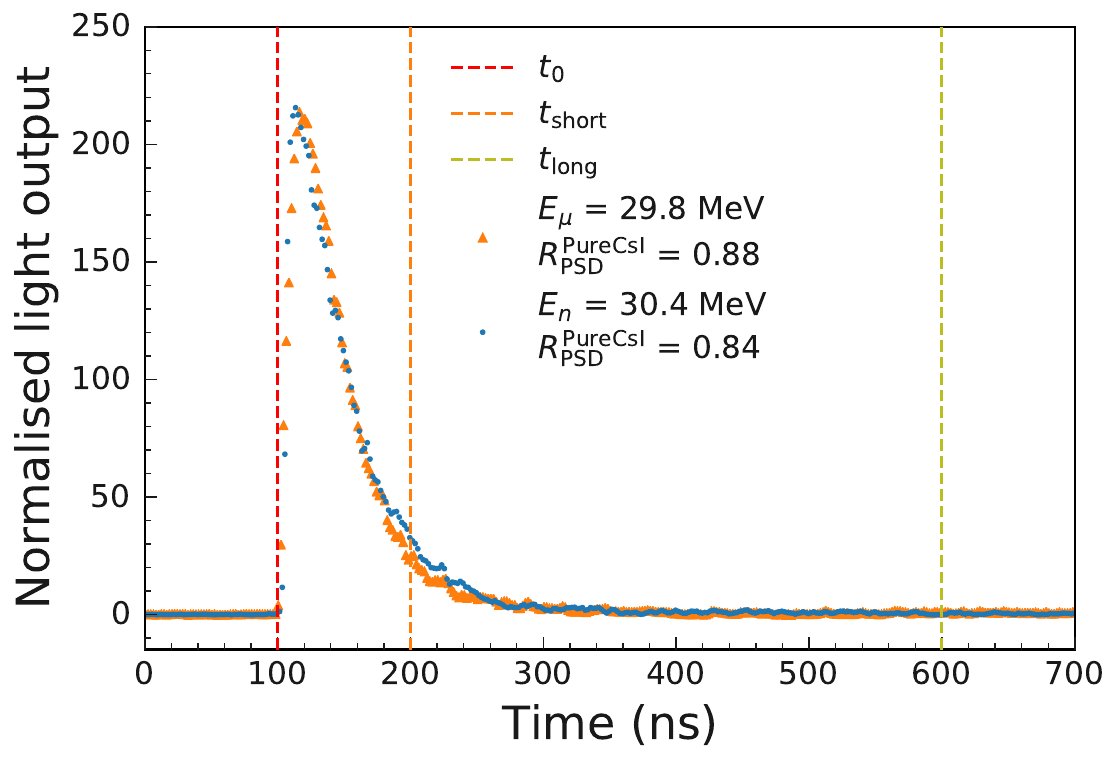}
\includegraphics[width=0.49\textwidth]{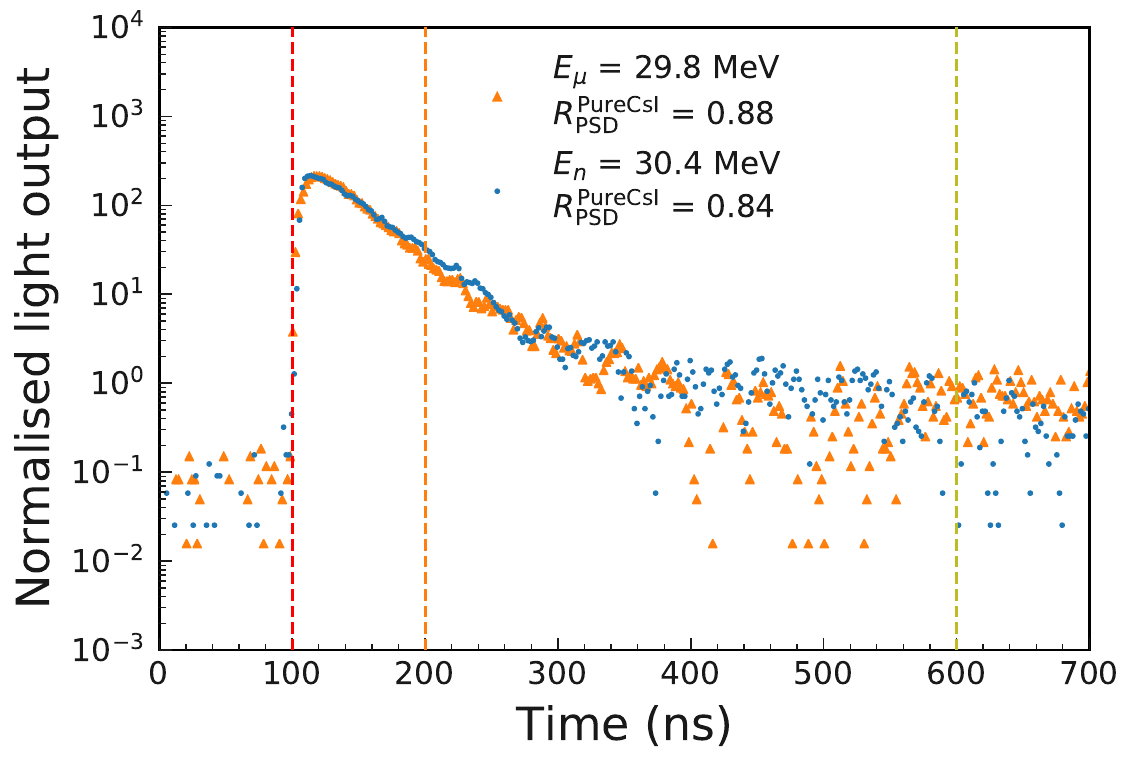}
  \caption{}
\end{subfigure}

\begin{subfigure}{1\textwidth}
  \centering
\includegraphics[width=0.49\textwidth]{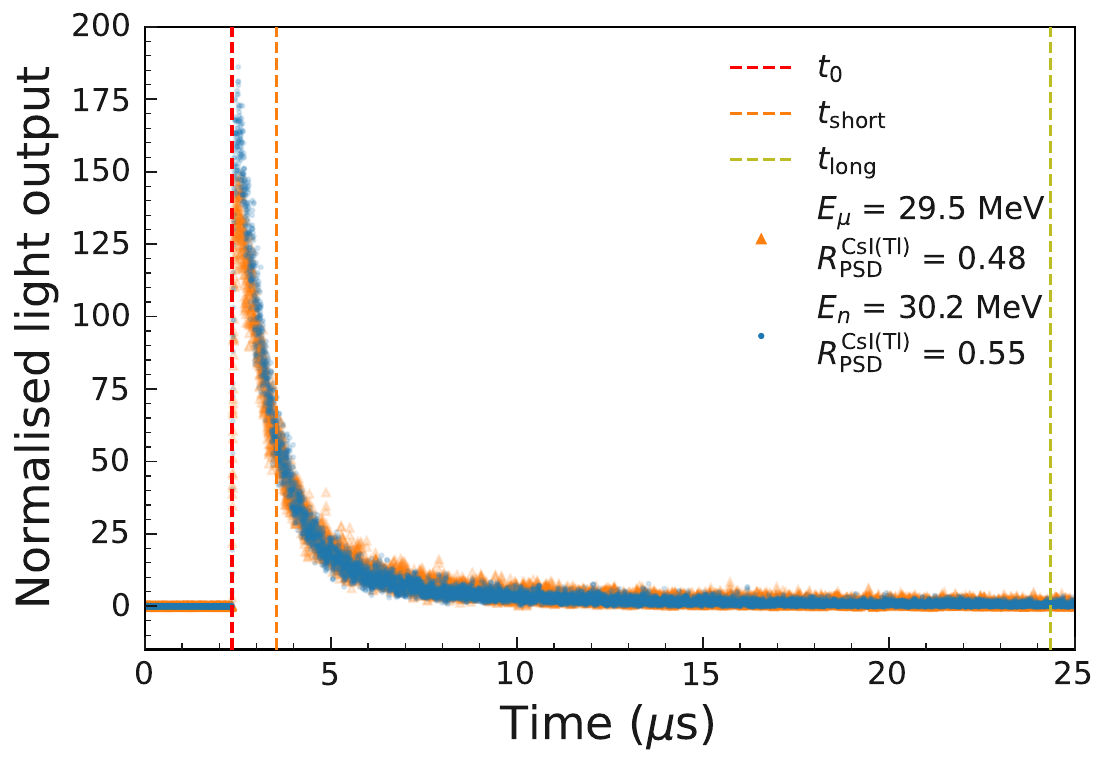}
\includegraphics[width=0.49\textwidth]{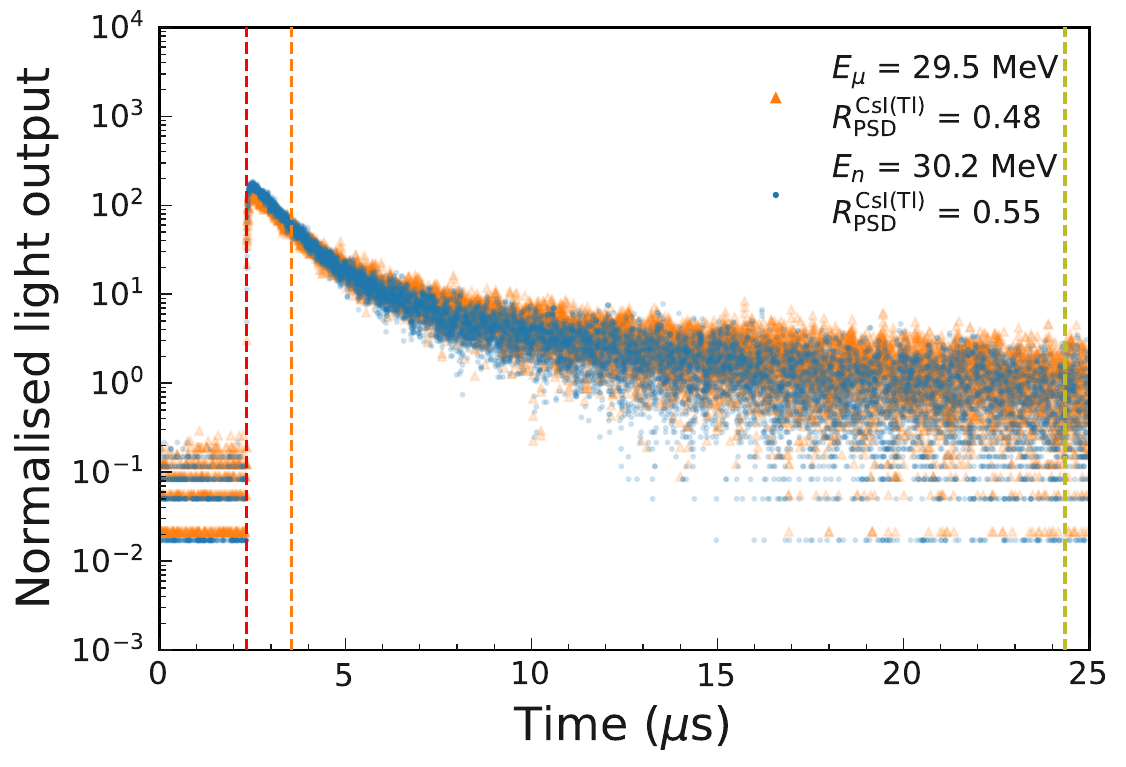}
  \caption{}
\end{subfigure}

\caption{Overlay of a typical muon-like and neutron-like waveform for a) pure CsI and b) CsI(Tl).  The long time window corresponds to the integration time used to compute the total energy.  Also overlaid is the short time window used for pulse shape characterization detailed as in Section \ref{sec_PSD}.}
\label{fig_typical_waveform}
\end{figure}

Shown in Figure \ref{fig_typical_waveform} is an overlay of a typical waveform with muon-like and neutron-like pulse shape for \purecsi and \csitl. The waveform $t_0$ is defined as the start of the scintillation emission and is measured using an Optimum Filter fit with the \texttt{QETpy} software framework \cite{caleb_w_fink_2021_5104856}. 
The energy deposited is computed by integrating the waveform over a long time window beginning from the waveform $t_0$. The long time windows applied correspond to $t_\mathrm{long}^\mathrm{pure} = 500$\,ns for \purecsi and  $t_\mathrm{long}^\mathrm{Tl} = 22\,\upmu$s for \csitl. These time windows are used as they contain the significant majority of the primary scintillation emission as shown in Figure \ref{fig_typical_waveform}. 
Energy calibration is completed off-site using the 0.569\,MeV, 1.063\,MeV and 1.770\,MeV primary photons emitted from a $^{207}$Bi source as well as 1.46 MeV photons emitted from natural $^{40}$K background \cite{Zyla:2020zbs,MuniraThesis}.  
The energy calibration stability was cross-checked on-site using the peaking energy distribution from cosmic muons.  
The charge-ratio pulse shape characterization presented in Section \ref{sec_PSD} is insensitive to the energy calibration.

\subsection{Energy deposited in beam-off and beam-on datasets}

\begin{figure}[ht]

\begin{subfigure}{.5\textwidth}
  \centering
\includegraphics[width=1\textwidth]{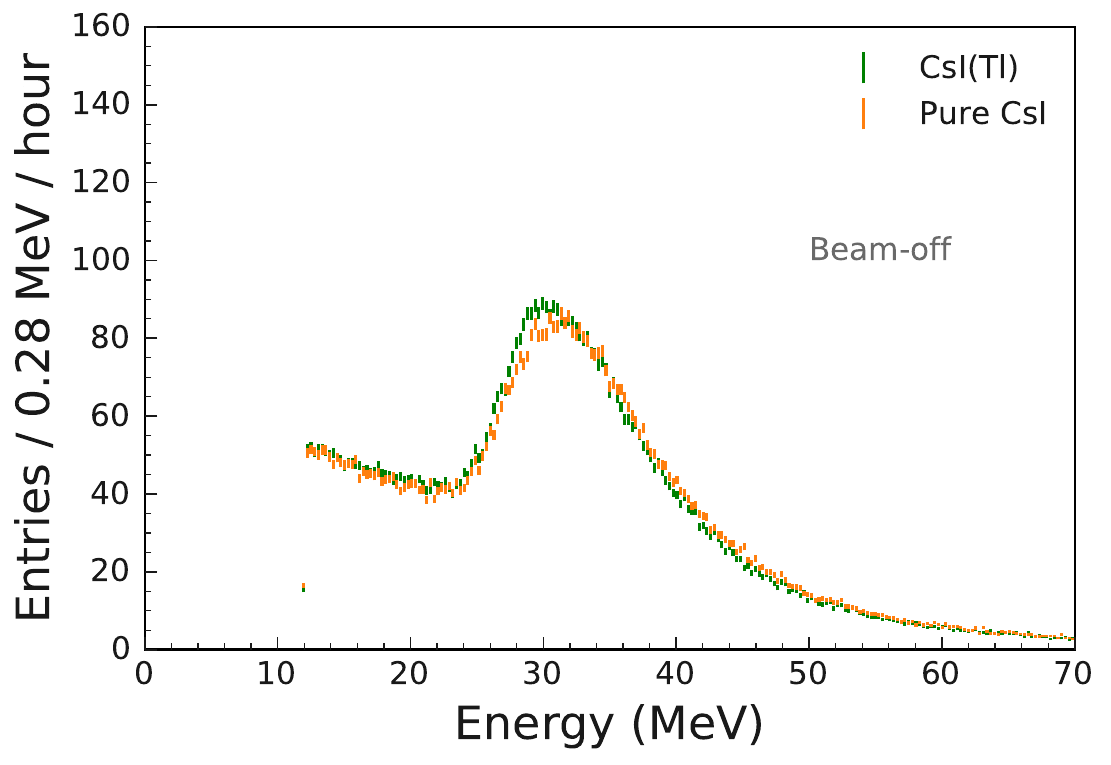}\\
  \caption{}
  \label{fig_EnergyHistograms_beamoff}
\end{subfigure}
\begin{subfigure}{.5\textwidth}
  \centering
\includegraphics[width=1\textwidth]{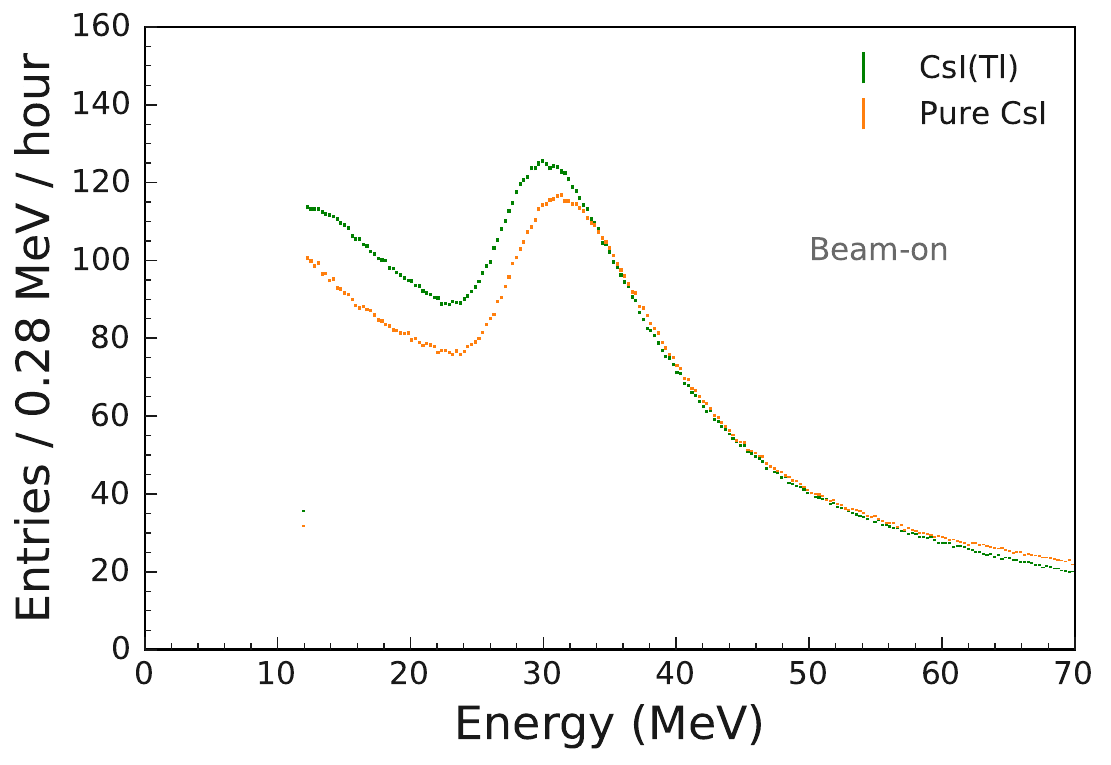}\\
  \caption{}
   \label{fig_EnergyHistograms_beamon}
\end{subfigure}

\centering
\begin{subfigure}{.5\textwidth}
  \includegraphics[width=1\linewidth]{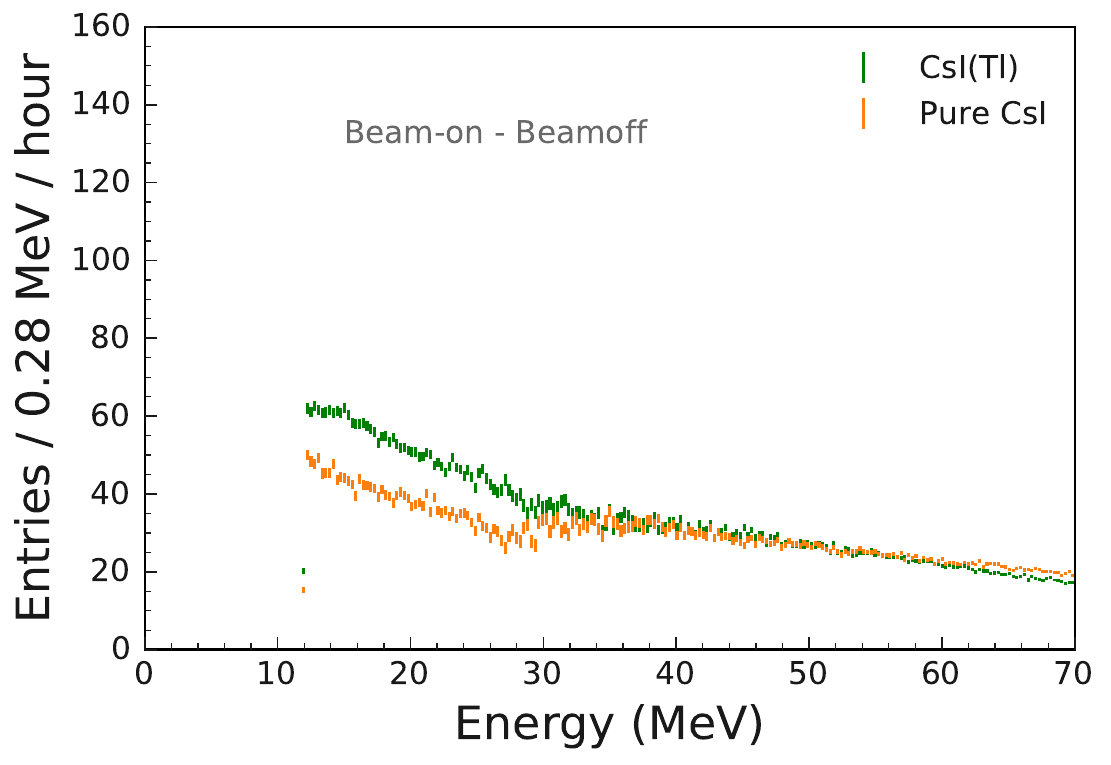}
  \caption{}
    \label{fig_EnergyHistograms_difference}
\end{subfigure}

\caption{Distribution of the waveform total energy for waveforms recorded in the a) beam-off and b) beam-on datasets after the pre-selection is applied.  Results for \purecsi and \csitl are overlaid for comparison.  The beam-on data corresponds to 491.75~hours of data-taking and the beam-off corresponds to 33.75~hours. Each distribution is normalized to show the hourly rate as a function of energy. In c) the energy deposits from neutrons are isolated by subtracting the beam-off histogram from the beam-on histogram. }
\label{fig_EnergyHistograms}
\end{figure}

The distribution of the measured energy for waveforms in the beam-off dataset is shown in Figure \ref{fig_EnergyHistograms_beamoff} with the pure CsI and CsI(Tl) results overlaid. The shape of this distribution is determined by the crystal geometry as it controls the track length of the muon ionization in the crystal.  As the crystals have near-identical geometries, the \purecsi and \csitl distribution are nearly identical in Figure \ref{fig_EnergyHistograms_beamoff}.  The peak at approximately 30\,MeV corresponds to the energy deposits from the most probable muon track length.  Although the energy resolution of \csitl is much better than \purecsi due to the larger light output yield, the resolution of the peak in Figure \ref{fig_EnergyHistograms_beamoff} is comparable for \purecsi and \csitl as the peak width is limited by the variations in the muons track length, which is the same for both crystals.

Figure \ref{fig_EnergyHistograms_beamon} shows the distribution of the energy deposited for waveforms in the beam-on dataset. The beam-on dataset corresponds to waveforms recorded while the \xfel beam was in operation and as a result the dataset contains energy deposits both from energetic neutron interactions and cosmic muons. The peak from cosmic muon ionization is still observed in Figure  \ref{fig_EnergyHistograms_beamon} due to the continuous flux of cosmic muons. In Figure \ref{fig_EnergyHistograms_difference} the beam-off distribution is subtracted from the beam-on distribution, allowing the distribution of energy deposited from the energetic neutrons to be observed. The rate of energetic neutrons is observed to be similar to the rate of cosmic muons. 

\section{Pulse Shape Discrimination with \Purecsi}
\label{sec_PSD}

\subsection{Pulse shape characterization}

The shape of the \purecsi and \csitl waveforms are characterized using a standard charge-ratio method \cite{DINCA2002141,Longo:2018uyj,knoll2010radiation}.
The charge-ratio is computed from the total scintillation light yield over a short time window, $E(t_\text{short})$, normalized to the total scintillation light yield over a long time window, $E(t_\text{long})$.  
Energy deposits that generate faster scintillation emission will result in a larger charge-ratio.  
The charge-ratios used for \purecsi and \csitl are defined as
\begin{equation}\label{eqn_pure_ratio}
R_{\rm PSD}^{\rm \text{\Purecsi}} \equiv \frac{E(100\,{\rm ns})}{E(500\,{\rm ns})},
\end{equation}

\begin{equation}\label{eqn_Tl_ratio}
R_{\rm PSD}^{\rm \csitl} \equiv \frac{E(1.2\,{\rm \upmu s})}{E(22\,{\rm \upmu s})}.
\end{equation}
The short time window of 1.2\,$\upmu$s is applied for \csitl as this has been reported to be optimal in previous studies of pulse shape discrimination in with \csitl \cite{MCLEAN2006793}. 
For \purecsi a short time window of 100\,ns is applied as it was determined to give the optimal neutron/muon separation, as detailed in Section \ref{sec_optimalChargeRatio}.  

\subsection{Comparison of pulse shape discrimination in \purecsi and \csitl}

\begin{figure}[ht]
\begin{subfigure}{.5\textwidth}
  \centering
  \includegraphics[width=1\linewidth]{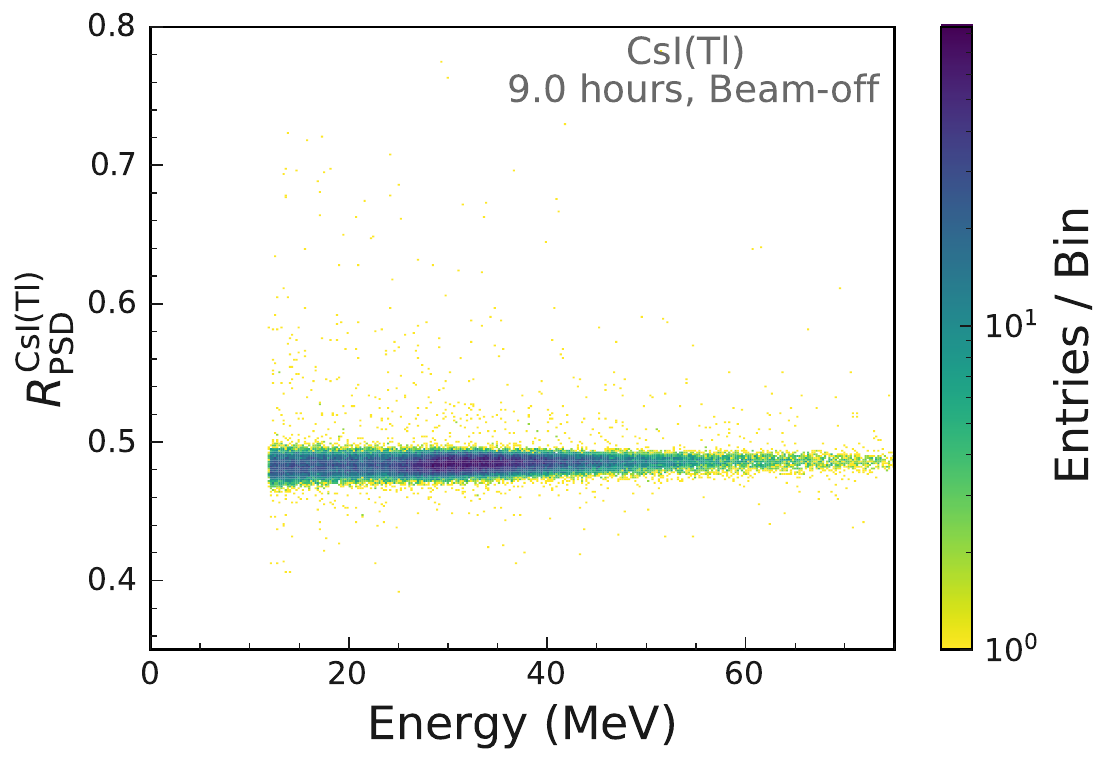}
  \caption{}
  \label{fig_Tl_beamoff}
\end{subfigure}
\begin{subfigure}{.5\textwidth}
  \centering
  \includegraphics[width=1\linewidth]{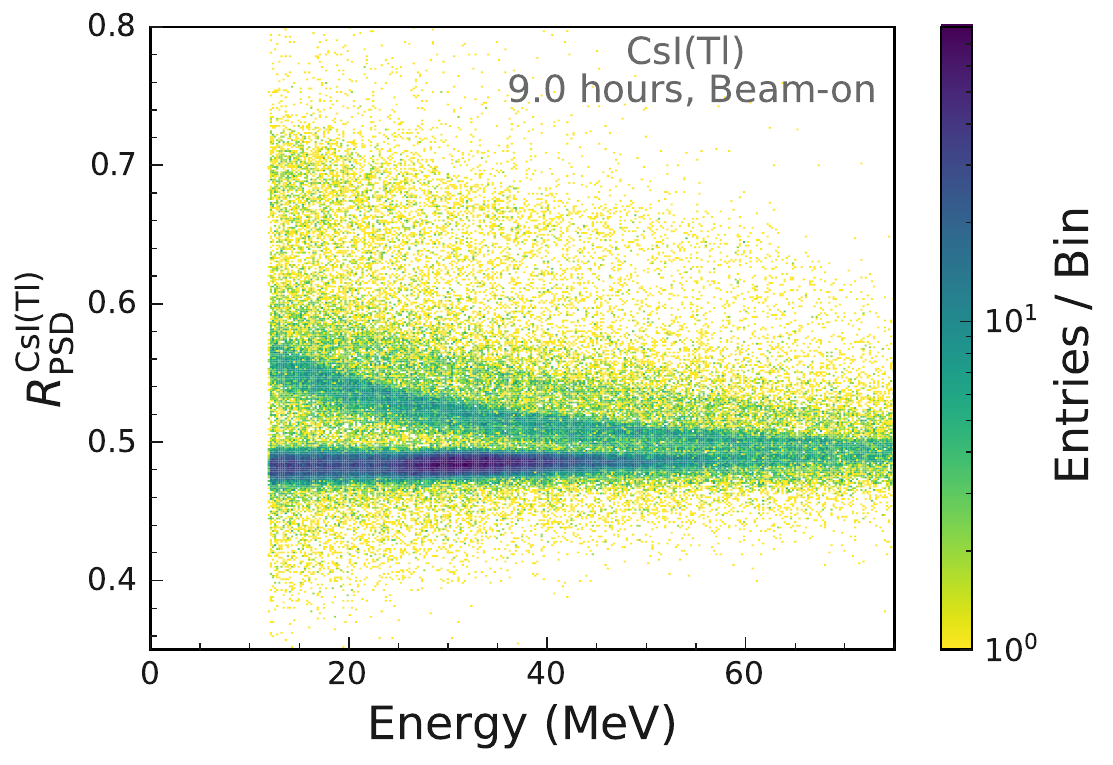}
  \caption{}
  \label{fig_Tl_beamon}
\end{subfigure}

\begin{subfigure}{.5\textwidth}
  \centering
  \includegraphics[width=1\linewidth]{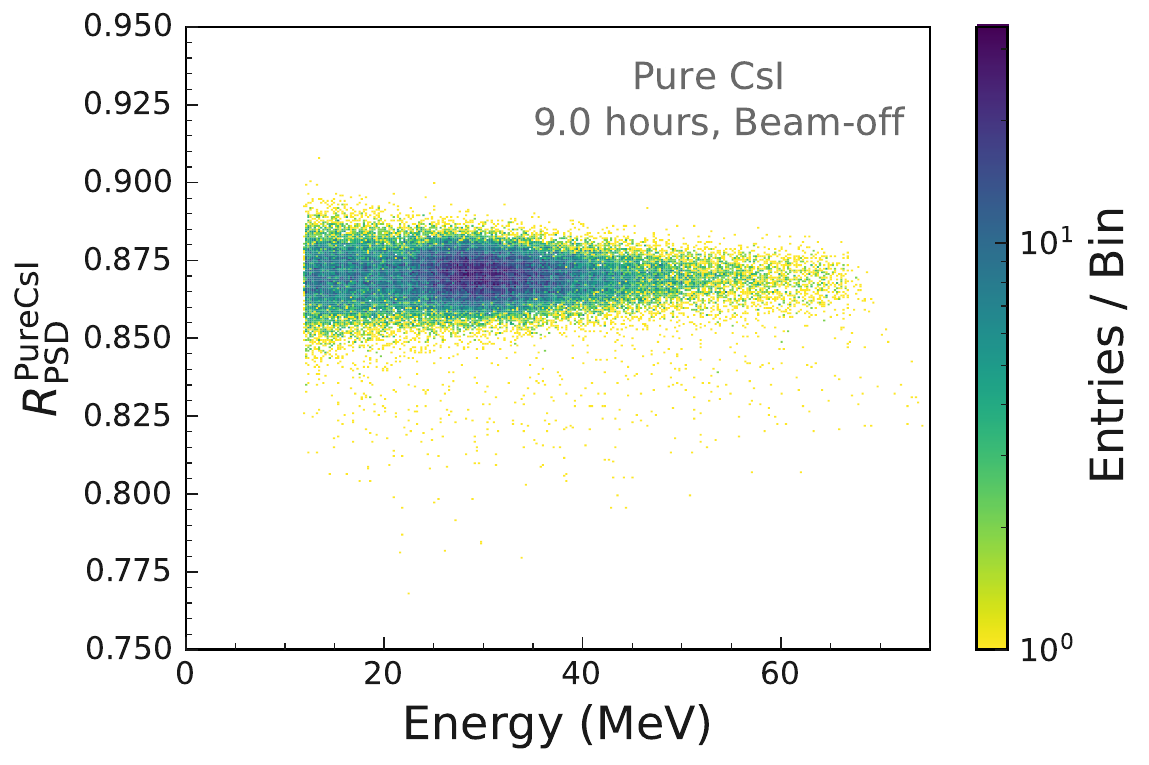}
  \caption{}
  \label{fig_pure_beamoff}
\end{subfigure}
\begin{subfigure}{.5\textwidth}
  \centering
  \includegraphics[width=1\linewidth]{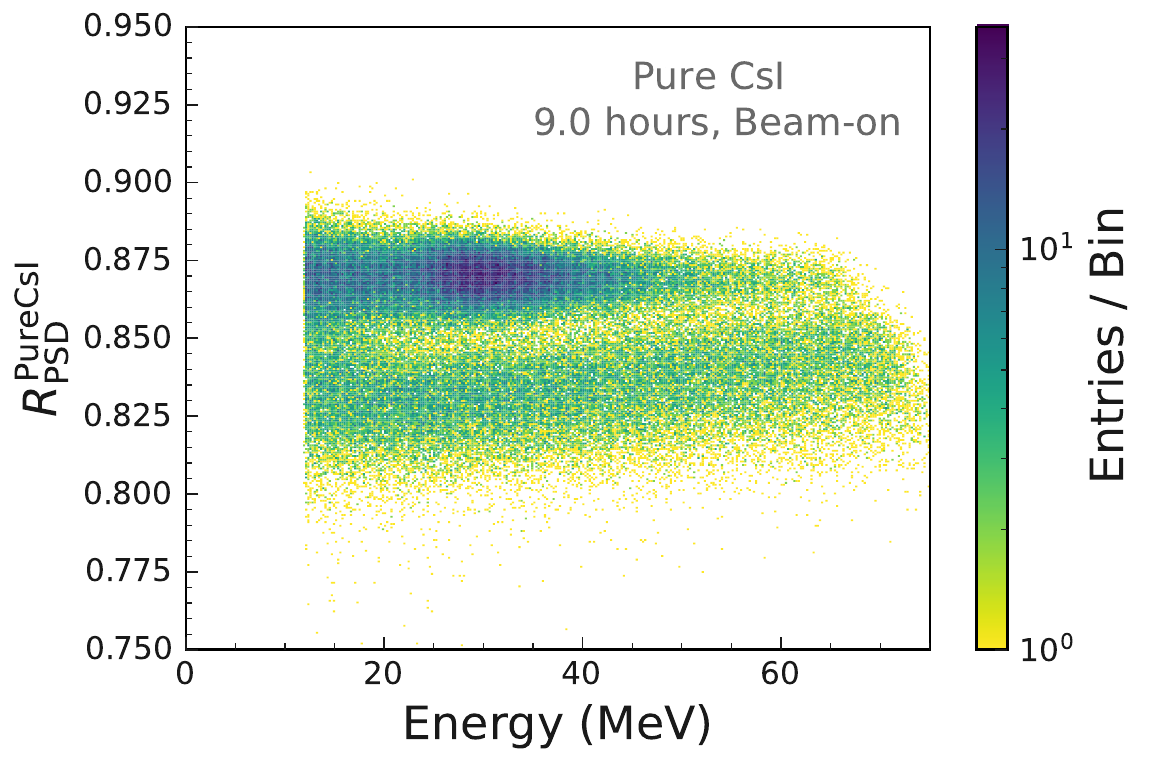}
\caption{}
  \label{fig_pure_beamon}
\end{subfigure}

\caption{Two-dimensional histogram of the charge-ratio vs. total energy for the waveforms in  a 9 hour time block for  a) \csitl beam-off, b) \csitl beam-on, c) \purecsi beam-off and d) \purecsi beam-on. In both the \purecsi and \csitl beam-on data an additional population of waveform shapes, which is not present in the beam-off data, is present and corresponds to energy deposits from neutron inelastic interactions in the crystal.  In the case of \csitl, the energy deposits from neutron interactions result in a faster scintillation time relative to the cosmic muons leading to larger charge-ratios.  The opposite trend is observed for \purecsi such that the neutron interactions result in lower charge-ratios relative to the muon energy deposits, indicating a slower scintillation time.
}
\label{fig_chargeRatio_plots}
\end{figure}

To illustrate the distribution of pulse shapes present in the beam-on and beam-off datasets, Figure \ref{fig_chargeRatio_plots} presents a two-dimensional histogram of the charge-ratio vs. total energy for the waveforms in a 9 hour time block of each dataset studied.  For both the beam-on and beam-off results shown in Figure \ref{fig_chargeRatio_plots}, the \csitl and \purecsi detectors were operated in parallel and during the same operational periods such that both detectors were exposed to the same radiation fluxes.  As described in Section \ref{sec_Experimentaldata}, cosmic muons are the primary source of the energy deposits in the beam-off dataset and in the beam-on dataset an additional flux of energetic neutrons is present leading to a dataset containing energy deposits from cosmic muons and neutrons.

Observing the beam-off \csitl data in Figure \ref{fig_Tl_beamoff}, the majority of waveforms have a charge-ratio distributed around 0.48, independent of the waveform energy.  This pulse shape behaviour has been observed in previous studies of the \csitl scintillation response to muons \cite{Longo:2018uyj,Longo:2020zqt}. In Figure \ref{fig_Tl_beamon} the beam-on \csitl data is shown.  In the beam-on data, multiple pulse shape features arise in the two-dimensional histogram at charge-ratios above 0.5. The band structures at larger charge-ratios have been observed in previous studies of the \csitl scintillation response to energetic neutron interactions \cite{Longo:2018uyj,BARTLE199954,MCLEAN2006793}.  As the \csitl scintillation time is faster for energy deposits by highly ionizing particles, the \csitl charge-ratio increases as the fraction of energy deposited by highly ionizing particles increases \cite{Longo:2018uyj}.
Each band in Figure \ref{fig_Tl_beamon} corresponds to a specific combination of highly ionizing secondary particles emitted from the neutron inelastic interaction in the \csitl. The two most prominent bands observed are from neutron inelastic interactions that emitted one proton and two protons, which then ionize and stop in the crystal.  The population of waveforms at 15\,MeV and charge-ratio of 0.7 is from neutron inelastic interactions that emitted an alpha particle that ionizes and stops in the crystal \cite{Longo:2018uyj}. 

In Figure \ref{fig_pure_beamoff} the distribution of pulse shapes for the waveforms in the \purecsi beam-off data is shown.  Similar to the \csitl beam-off data, a single band of constant charge-ratio is observed indicating the pulse shape is independent of the muon ionization track length.  In Figure \ref{fig_pure_beamon} the distribution of pulse shapes for the waveforms in the beam-on \purecsi data are shown.  Comparing Figures \ref{fig_pure_beamon} and \ref{fig_pure_beamoff}, a prominent second band is observed in Figure \ref{fig_pure_beamon} at a charge-ratio lower than the charge-ratio of the waveforms from the muon energy deposits.  The lower band in Figure \ref{fig_pure_beamon} is interpreted to be from neutron inelastic interactions in the crystal. Although the detectors triggered independently, the \purecsi and \csitl data are recorded during the same \xfel operational periods and as a result both crystals are exposed, on average, to the same neutron flux.  As the \csitl and \purecsi crystals have the same dimensions, the distribution of the types of neutron interactions that occurred in each crystal is expected the be the same for \csitl and \purecsi when averaged over the entire beam-on dataset. The observation of the band in the \purecsi beam-on data with charge-ratio lower than the band from muon ionization indicates the scintillation emission in \purecsi for energy deposits from the inelastic neutron interactions is slower relative to the muon ionization.  This trend differs from \csitl where a faster scintillation emission is observed for highly ionizing particles. This observation is similar to the results presented by the KOTO experiment, where it was shown that after applying shaping electronics the resulting detector response to neutrons is slower compared to photons \cite{Sugiyama:2021ltp}.  Another interesting observation is that in the \purecsi beam-on data only a single band from neutron interactions is resolved.  This differs from \csitl where multiple bands are resolved allowing the specific secondaries emitted from the neutron interactions to be identified.

\begin{figure}[H]
\begin{subfigure}{.5\textwidth}
  \centering
  \includegraphics[width=1\linewidth]{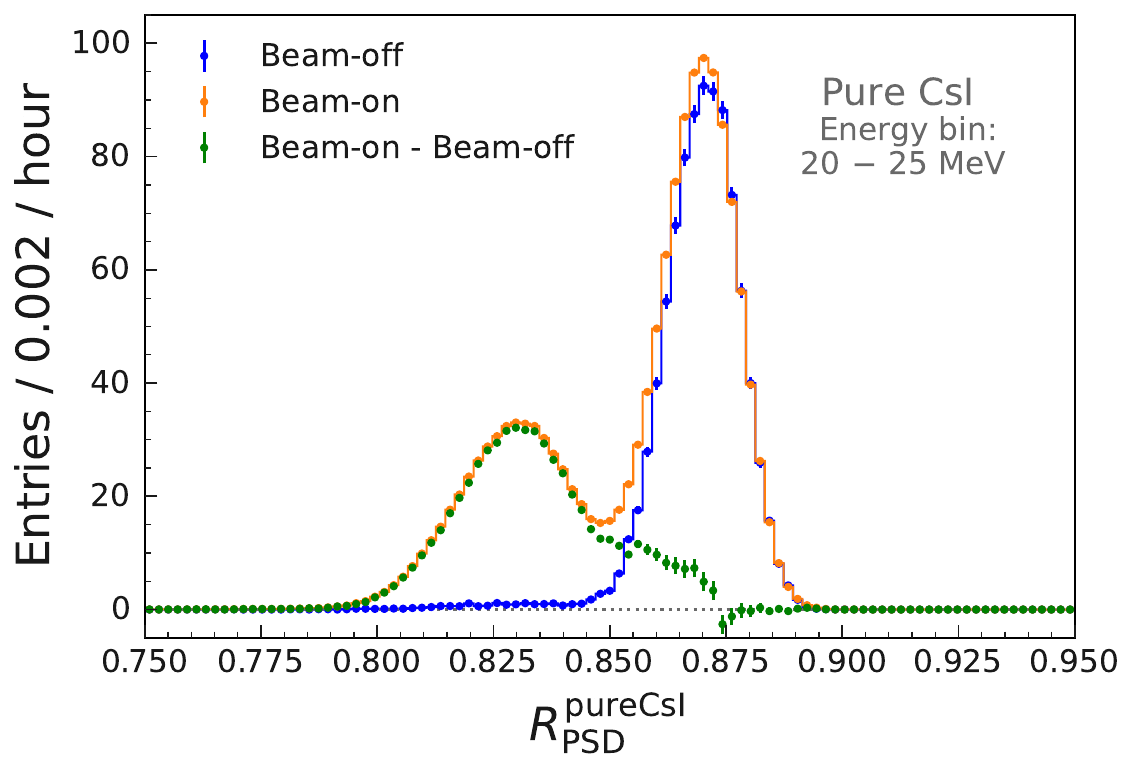}
  \caption{}
\end{subfigure}
\begin{subfigure}{.5\textwidth}
  \centering
  \includegraphics[width=1\linewidth]{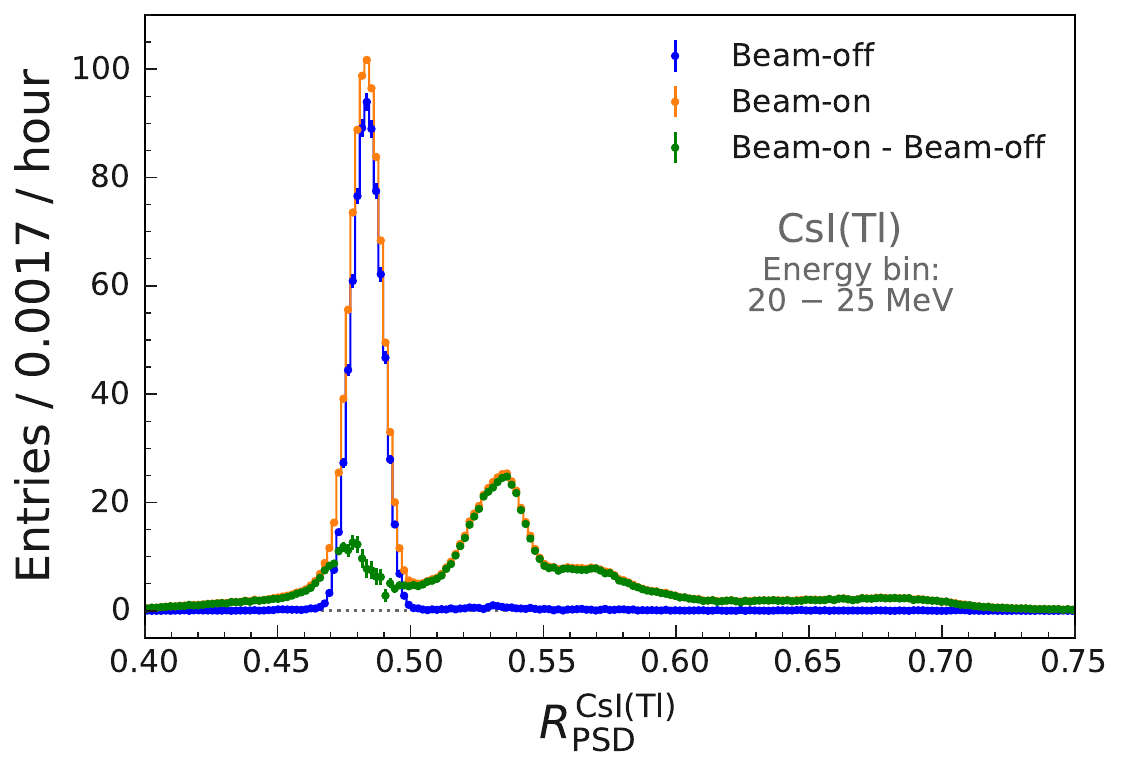}
  \caption{}
\end{subfigure}

\begin{subfigure}{.5\textwidth}
  \centering
  \includegraphics[width=1\linewidth]{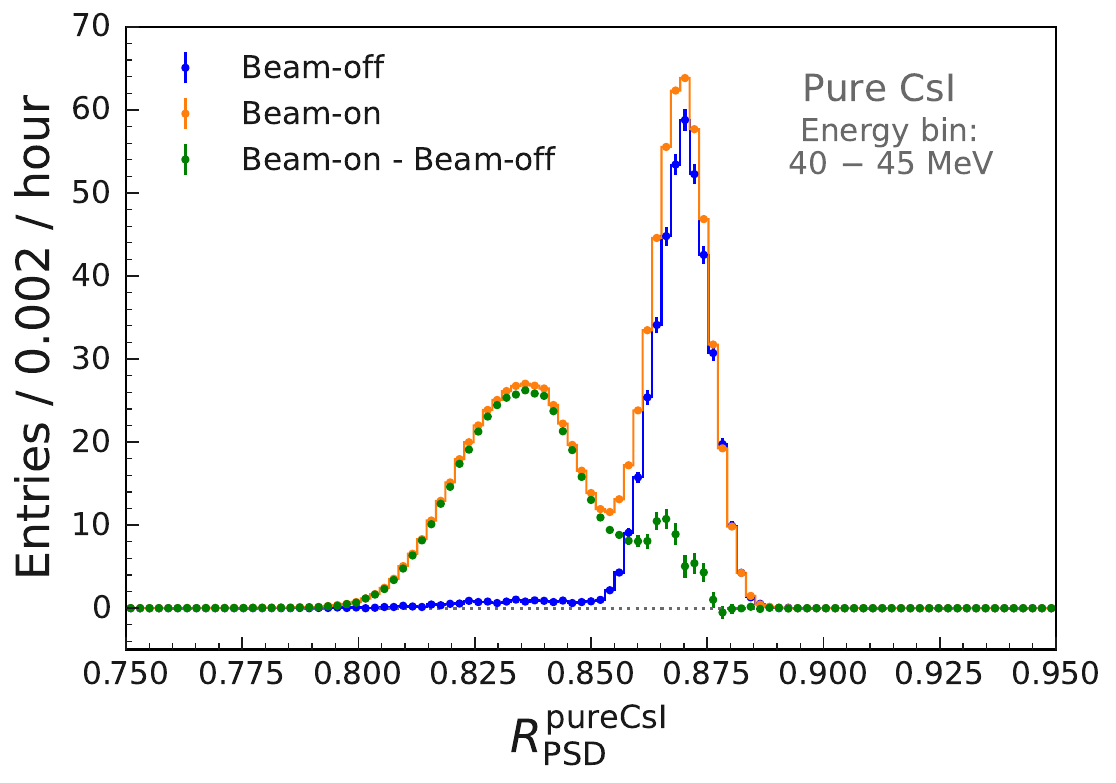}
  \caption{}
\end{subfigure}
\begin{subfigure}{.5\textwidth}
  \centering
  \includegraphics[width=1\linewidth]{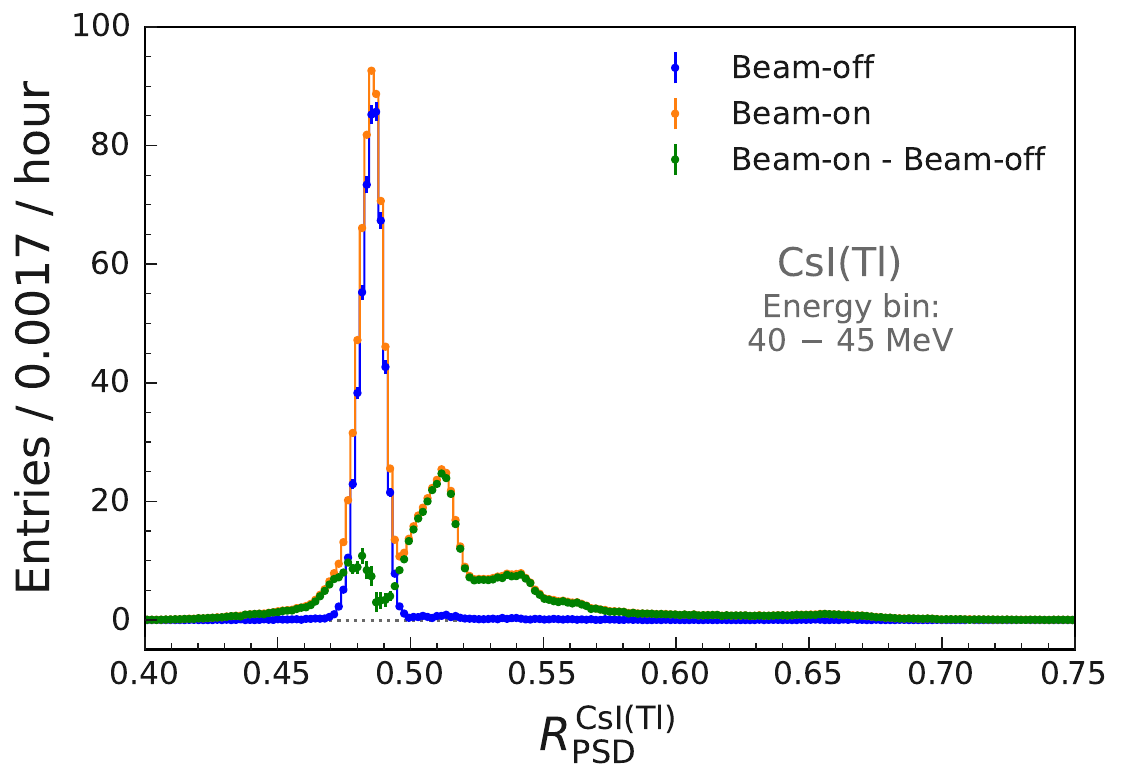}
  \caption{}
\end{subfigure}

\caption{Typical results for the charge-ratio distribution of waveforms in the beam-on and beam-off datasets for two of the energy bins studied.  The Beam-on\,--\,Beam-off distribution allows the charge-ratio distribution for the neutron interactions to be isolated.}

\label{fig_chargeRatio_1DHistos}
\end{figure}

In Figure \ref{fig_chargeRatio_1DHistos} one-dimensional charge-ratio histograms are shown for two 5\,MeV wide energy bins for \purecsi and \csitl.  Overlaid in each figure is the charge-ratio distribution for the beam-off and beam-on datasets in addition to the background subtracted beam-on data, computed by subtracting the beam-off distribution from the beam-on distribution.  The subtracted distributions in Figure \ref{fig_chargeRatio_1DHistos} allow the charge-ratio distribution for the energy deposits from the neutron interactions to be isolated.  An interesting observation from the subtracted results in Figure \ref{fig_chargeRatio_1DHistos} is that for both \purecsi and \csitl the neutron energy deposits result in two general populations of pulse shapes, one that is well separated from the muon peak and another that is at a similar charge-ratio to the muon energy deposits.  As discussed above, the features at larger charge-ratio in the \csitl subtracted distribution arise from specific combinations of highly ionizing secondary particles emitted from the neutron inelastic interaction.  The peak in the \csitl-subtracted distribution at the lowest charge-ratio, which overlaps with the muon peak, is expected to arise from neutron inelastic interactions where an energetic proton is emitted and escapes the crystal volume before it becomes highly ionizing \cite{Longo:2018uyj}.  As the proton is not highly ionizing, the energy deposit in this case produces a pulse shape that is similar to muons.  Another interesting observation from Figure \ref{fig_chargeRatio_1DHistos} is that an analogous peak is observed in the \purecsi subtracted distribution.  For both the \purecsi and \csitl, approximately 20\% of neutron interactions resulted in a pulse shape in this peak independent of the waveform energy.  As the \purecsi and \csitl crystals have the same dimensions and are exposed to the same radiation environments, it is expected that the origin of this sample of energy deposits corresponds to the same type of neutron interactions occurring independently in each crystal. 

\subsection{Optimal charge-ratio for \purecsi}
\label{sec_optimalChargeRatio}

\begin{figure}[ht]
\begin{subfigure}{.5\textwidth}
  \centering
  \includegraphics[width=1\linewidth]{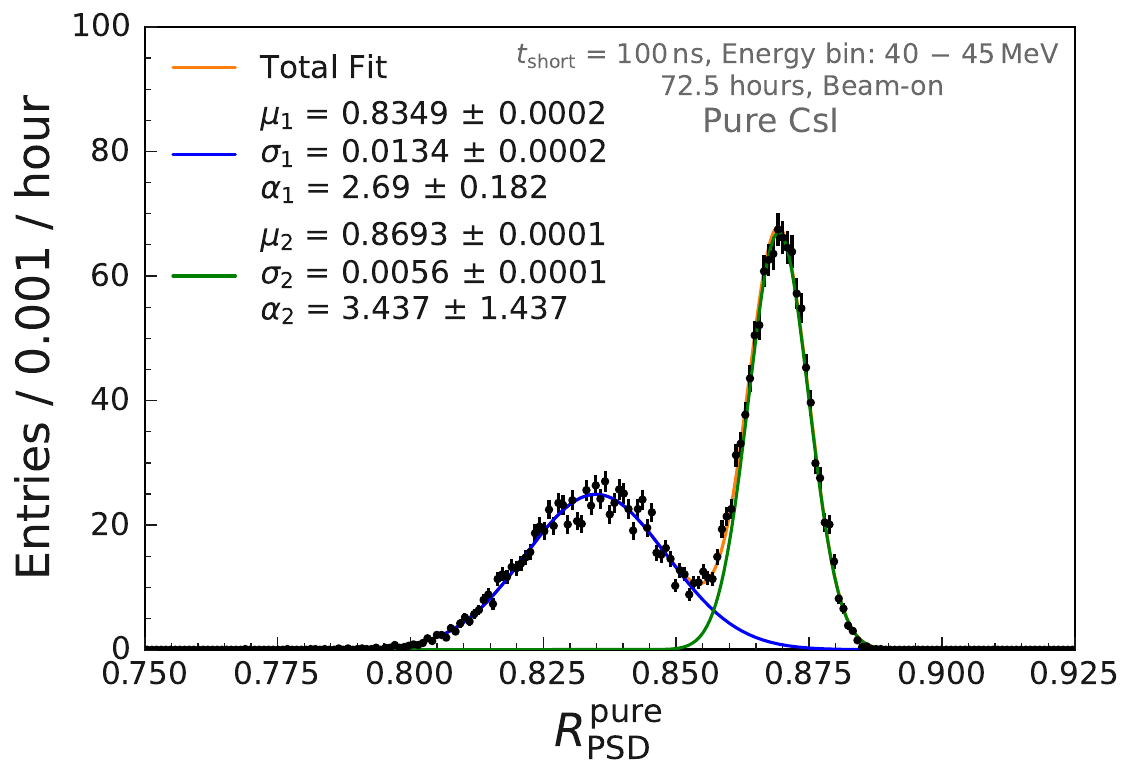}
  \caption{}
  \label{fig_sepPowerSummaryAna_fit}
\end{subfigure}
\begin{subfigure}{.5\textwidth}
  \centering
  \includegraphics[width=1\linewidth]{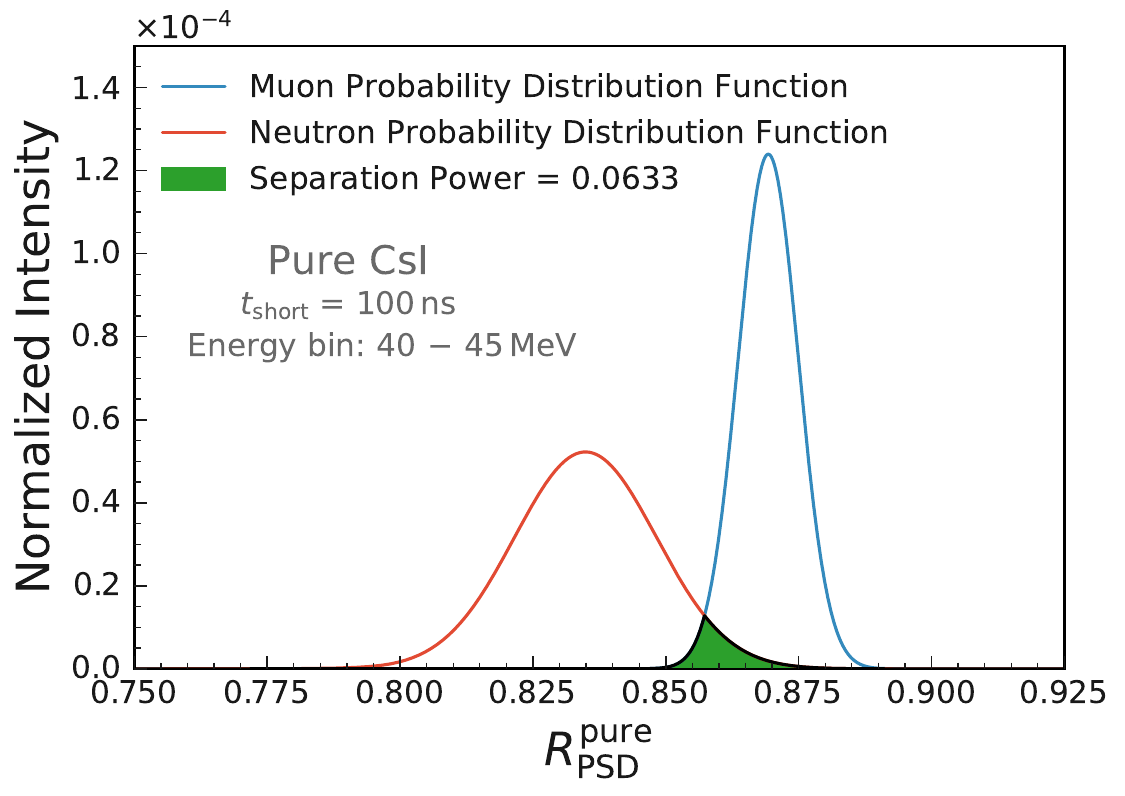}
  \caption{}
  \label{fig_sepPowerSummaryAna_eg}
\end{subfigure}

\caption{a) Typical fit to the beam-on charge-ratio distribution for the 40 - 45 MeV energy bin and short time of 100\,ns.  Both peaks are modelled by one-sided crystal ball functions.  b) Illustration of the separation power quantified for the distribution shown in a).  Each one-sided crystal ball distribution is normalized to unit area before the overlap region shown in green is computed. }

\label{fig_sepPowerSummaryAna}
\end{figure}

As illustrated in Figure \ref{fig_chargeRatio_1DHistos}, two distinct populations of waveform shapes are present in the beam-on \purecsi data. 
The population with charge-ratio around 0.87 is attributed to ionizing muons and the population at lower charge-ratio to neutron inelastic interactions. 
To determine the optimal \purecsi charge-ratio to separate these two populations the beam-on sample is divided into 5\,MeV wide energy bins from 25\,MeV to 55\,MeV and the overlap area of the neutron and muon distributions is computed as a function of the short time window. 
For each energy bin and short time window combination a one-dimensional histogram is produced that features a bimodal distribution as shown in Figure \ref{fig_sepPowerSummaryAna_fit}. 
Two Crystal Ball functions (CB) are fit to each histogram, such that one CB models the peak from muon ionization and the other models the peak from the neutron interactions.
As the yield of each CB is controlled by the properties of the muon and neutron flux, each CB is normalized to unity before the overlap integral of the two CB functions is computed.
The typical results for this procedure is demonstrated in Figure \ref{fig_sepPowerSummaryAna_eg} showing the analysis performed on the working point corresponding to the 40 - 45\,MeV energy bin and the short time window of 100\,ns. 
The area of the highlighted green region corresponds to the measured overlap area, which should be minimized in order to maximize the separation power. 

\begin{figure}[ht]
\centering
 \includegraphics[width=0.65\linewidth]{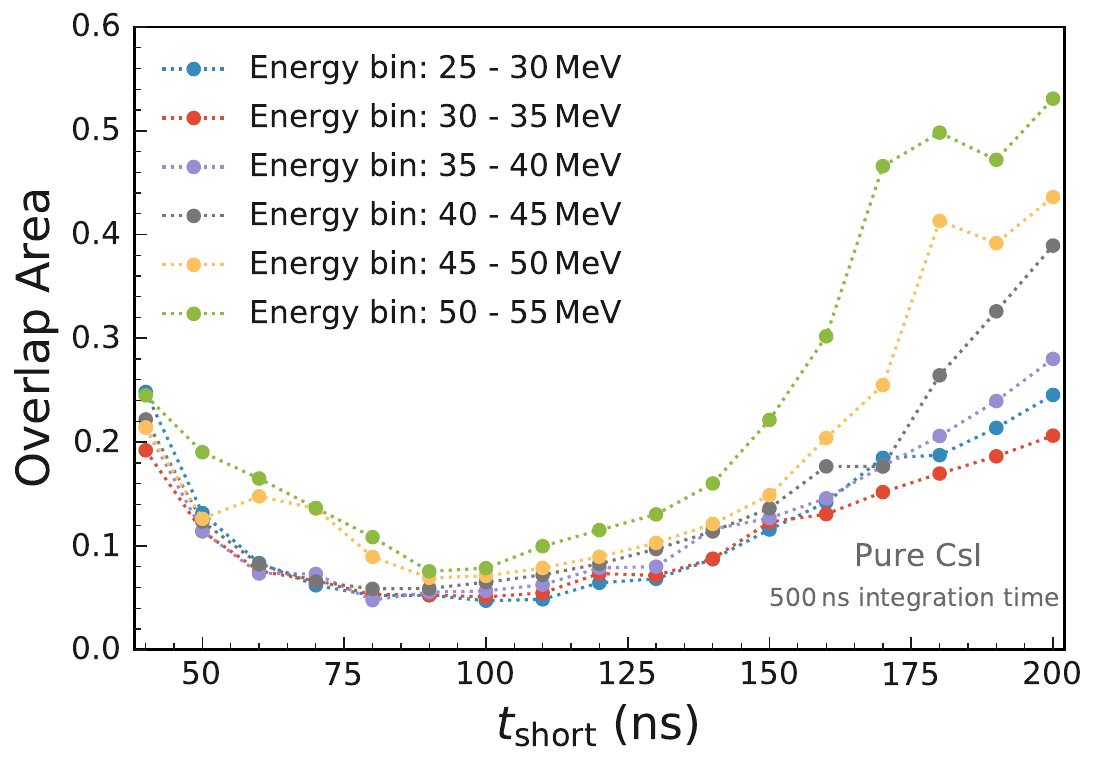}

\caption{Results of the short time optimization for \purecsi showing the measured separation power as a function of short time window for different bins in energy.  The short time of $100\pm10$\,ns is observed to be near optimal for all energy ranges studied.}

\label{fig_shortgateOptimization}
\end{figure}

The results of the short time window optimization are summarized in Figure \ref{fig_shortgateOptimization} showing the measured \purecsi overlap area as a function of the short time window for the multiple energy bins studied.  The short time window of $100\pm10$\,ns is observed to be near optimal for each energy bin. The overlap area also has only a weak dependence on the short time window between 75\,ns and 125\,ns.

\subsection{Energy dependence of \purecsi pulse shape and separation power}

\begin{figure}[ht]
\begin{subfigure}{.5\textwidth}
  \centering
  \includegraphics[width=1\linewidth]{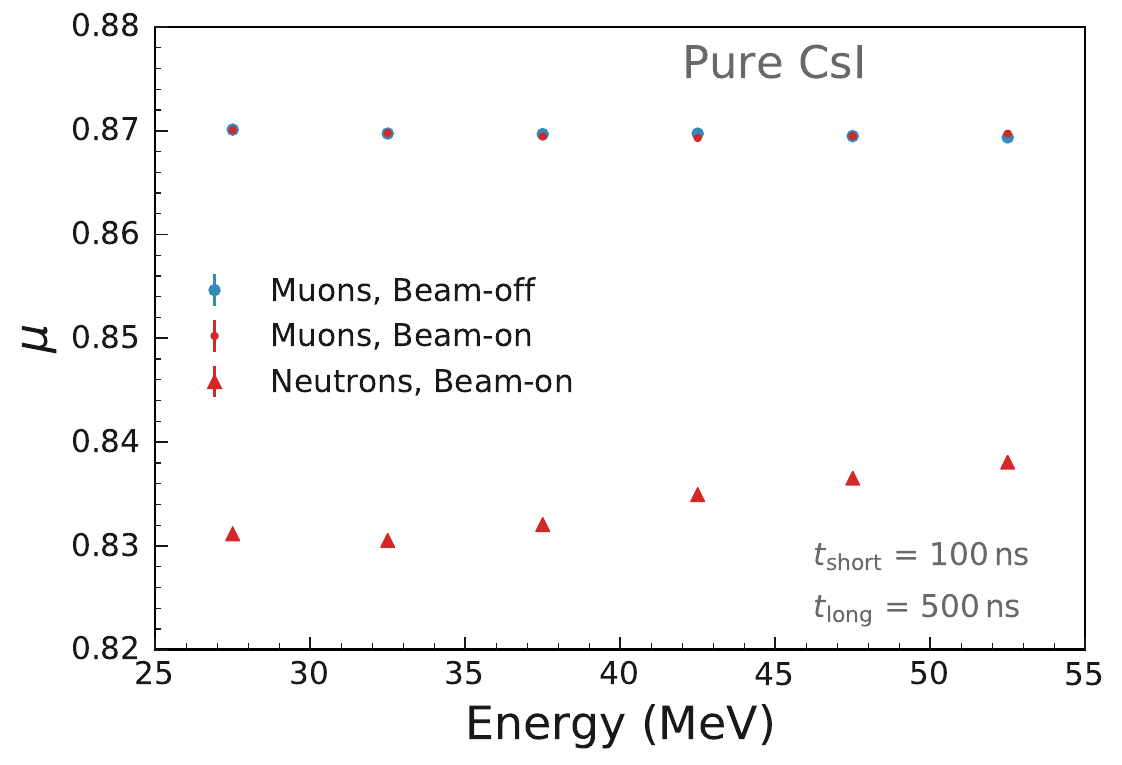}
  \caption{}
  \label{energydependence_mean}
\end{subfigure}
\begin{subfigure}{.5\textwidth}
  \centering
  \includegraphics[width=1\linewidth]{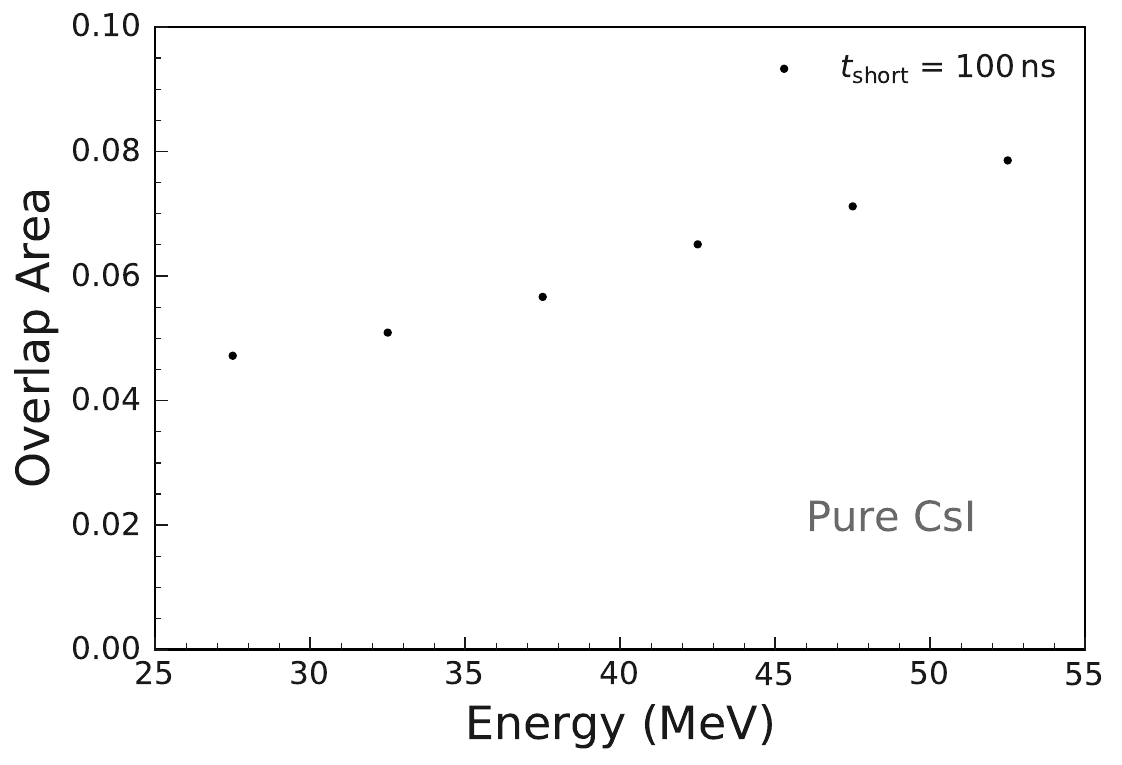}
  \caption{}
  \label{energydependence_operlaparea}
\end{subfigure}

\caption{a) Mean value of the crystal ball functions used to model the muon and neutron peaks shown as a function of the waveform energy. b) Overlap area between the muon and neutron probability distribution functions as a function of the energy deposited.}

\label{fig_energydependence}
\end{figure}

Figure \ref{energydependence_mean} shows the CB mean parameter, $\mu$, as a function of energy bin for the muon and neutron peaks observed in the \purecsi data.   Overlaid in this figure are the CB mean for the muon and neutron peaks in the beam-on data, as well as the muon peak in the beam-off data.  The mean values of the muon peak in the two datasets are in agreement as expected.  The muon peak is nearly constant as a function of energy with a mean value of 0.87.    The mean value of the neutron peak is observed to increase as the energy deposited increases.  In Figure \ref{energydependence_operlaparea} the overlap area is shown as a function of energy for \purecsi.  As the energy increases the overlap area is observed to increase leading to a reduction in the separation power. 

\section{Neutron and photon identification with \purecsi}
\label{sec_monitoring}

The pulse shape discrimination capabilities of pure CsI shown in the previous section demonstrate it can be an effective detector to isolate the rate of fast neutrons while also being irradiated by a background of energetic photons.  Using the muon and neutron probability distribution functions computed in the previous section, the energy-dependent charge-ratio threshold corresponding to a 99\% muon rejection rate in pure CsI is computed.  A waveform with charge-ratio below the 99\% threshold is classified as a neutron.  In 15 minute bins the number of neutrons was measured for several days of \xfel operation, during which the \xfel beams operated at a variety of different beam intensities, resulting in varying neutron rates.  

\begin{figure}[H]
  \centering
\begin{subfigure}{.75\textwidth}
  \centering
  \includegraphics[width=1\linewidth]{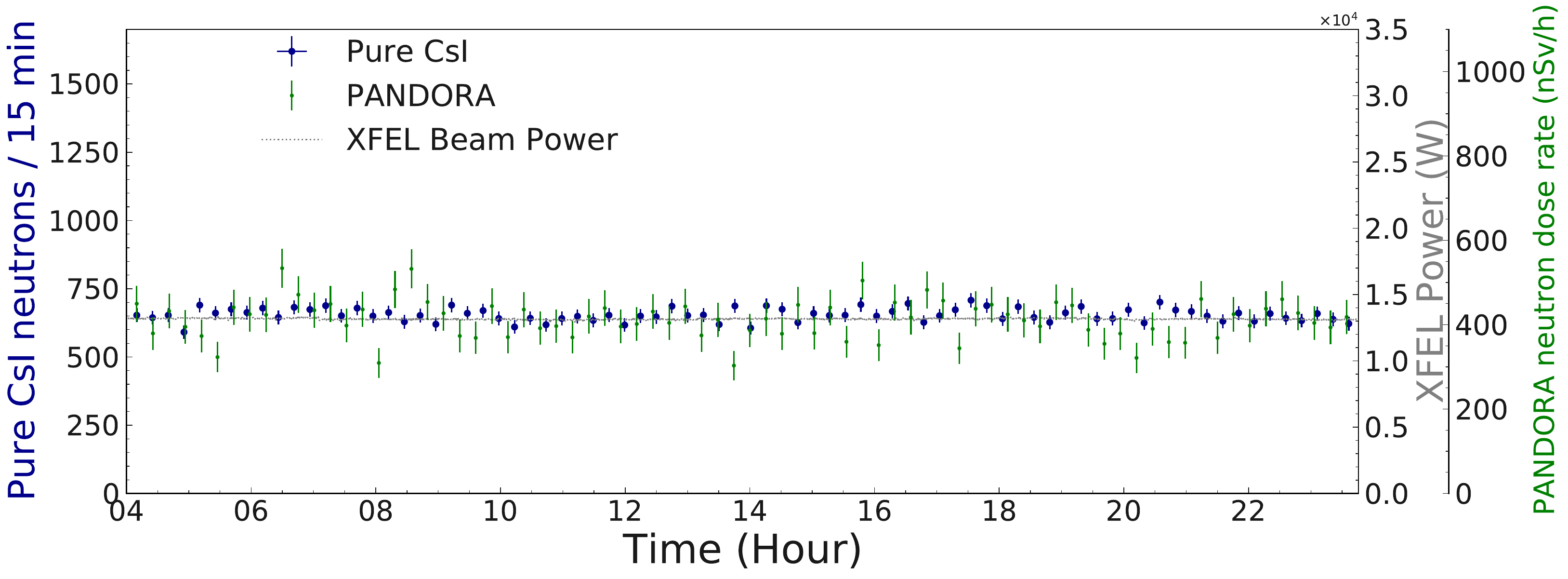}
  \caption{}
  \label{fig_neutronMonitor_stable}
\end{subfigure}

\begin{subfigure}{.75\textwidth}
  \centering
  \includegraphics[width=1\linewidth]{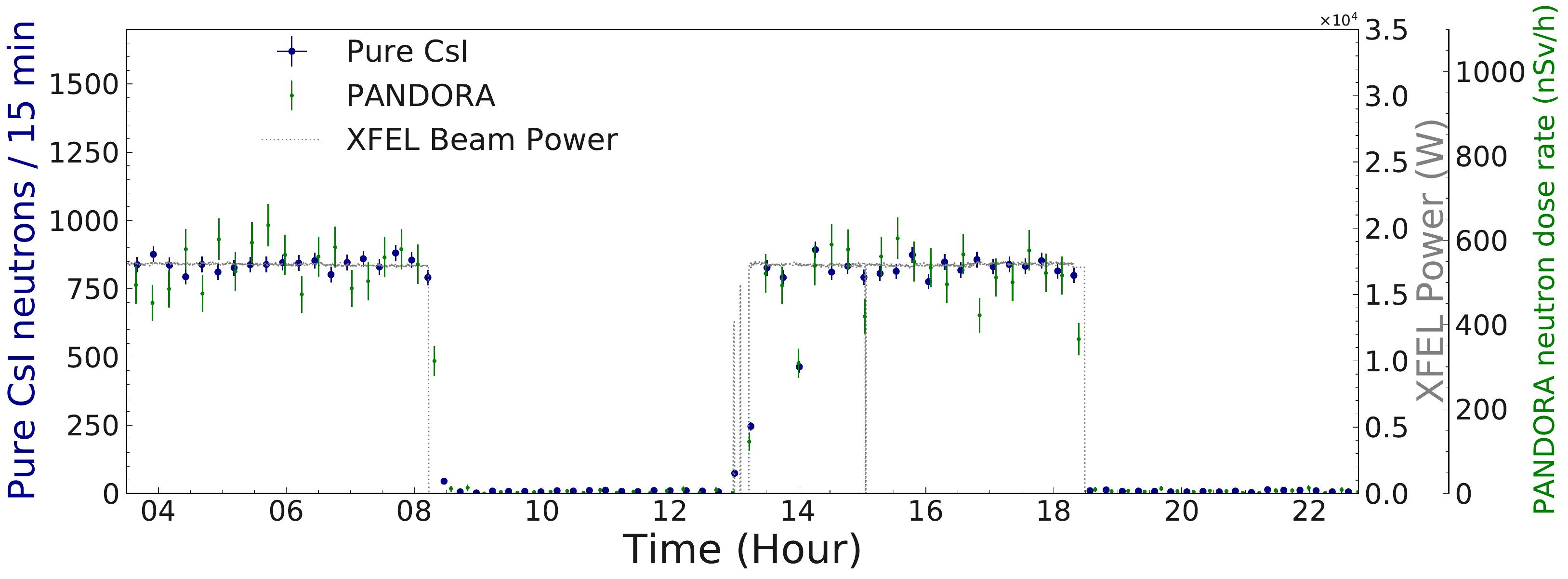}
  \caption{}
  \label{fig_neutronMonitor_offon}
\end{subfigure}

\begin{subfigure}{.75\textwidth}
  \centering
  \includegraphics[width=1\linewidth]{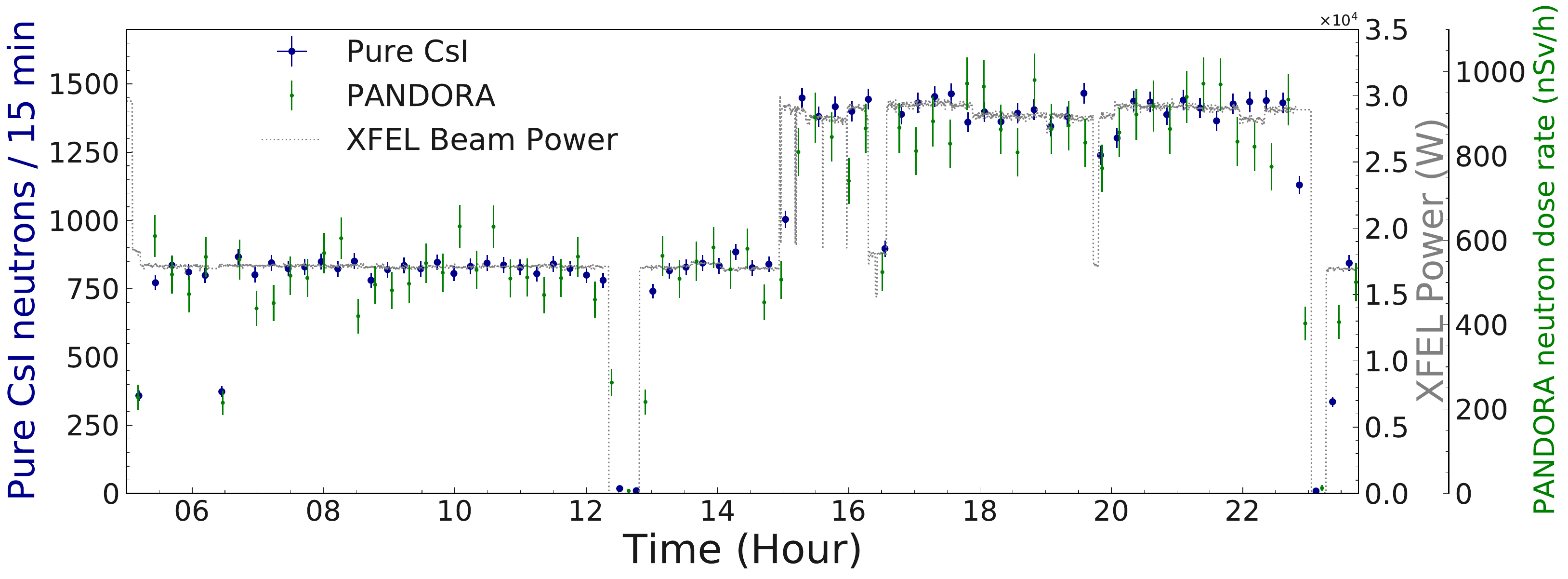}
  \caption{}
  \label{fig_neutronMonitor_short1}
\end{subfigure}

\begin{subfigure}{.75\textwidth}
  \centering
  \includegraphics[width=1\linewidth]{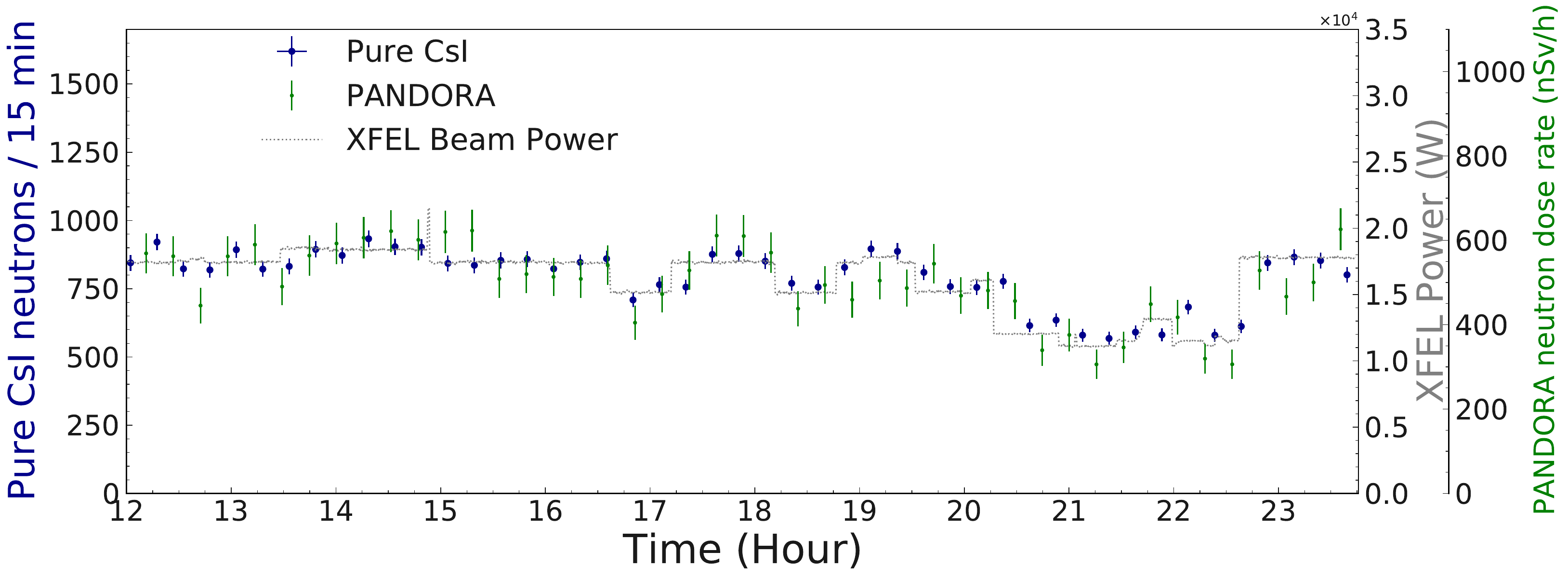}
  \caption{}
  \label{fig_neutronMonitor_short2}
\end{subfigure}
\caption{
Neutron rates as measured by using pulse shape discrimination with \purecsi and by \pandora, compared to the \xfel power. Error bars for the \purecsi and \pandora results correspond to statistical errors.  The absolute dose measurement by \pandora has a 10\% systematic error, which is not included in the error bars in the plot. The overall vertical axis scaling for the pure CsI neutron rate, \pandora energetic neutron dose, and \xfel beam power are set to allow for the comparison between the three quantities, which are expected to be correlated. }

\label{fig_neutronMonitor}
\end{figure}

Figure \ref{fig_neutronMonitor} presents the measured neutron rates in the \purecsi detector for several \xfel operation periods.  In Figure \ref{fig_neutronMonitor} the pure CsI data corresponds to the number of neutrons detected in the pure CsI crystal in the 15 minute time window.  For comparison, the high energy neutron dose rate recorded by the \pandora radiation detector system, described in Section \ref{sec2_ExpSetup}, is also overlaid.  Each figure additionally shows the \xfel beam power as a function of time.  

Observed in Figure \ref{fig_neutronMonitor}, the neutron rate recorded by the pure CsI detector closely follows the trends of the XFEL beam power.  In Figure \ref{fig_neutronMonitor_stable} the beam power was stable for several hours resulting in a stable neutron rate observed by the pure CsI detector.  Comparing the pure CsI rate to the \pandora dose rate, the pure CsI detector is shown to have less bin-to-bin fluctuations.  Figure \ref{fig_neutronMonitor_offon} shows the monitoring results for an XFEL operational period when the beam power fluctuated off and on.  Although the cosmic muon flux is continuously irradiating the pure CsI crystal, the neutron rate in the pure CsI detector drops to around 1\% when the beam is off.  This result demonstrates that the pure CsI detector with pulse shape discrimination can reject photon backgrounds while maintaining efficient neutron detection. Note this fake-rate is controlled by the charge-ratio threshold that is

\begin{figure}[H]
  \centering

\begin{subfigure}{.75\textwidth}
  \centering
  \includegraphics[width=1\linewidth]{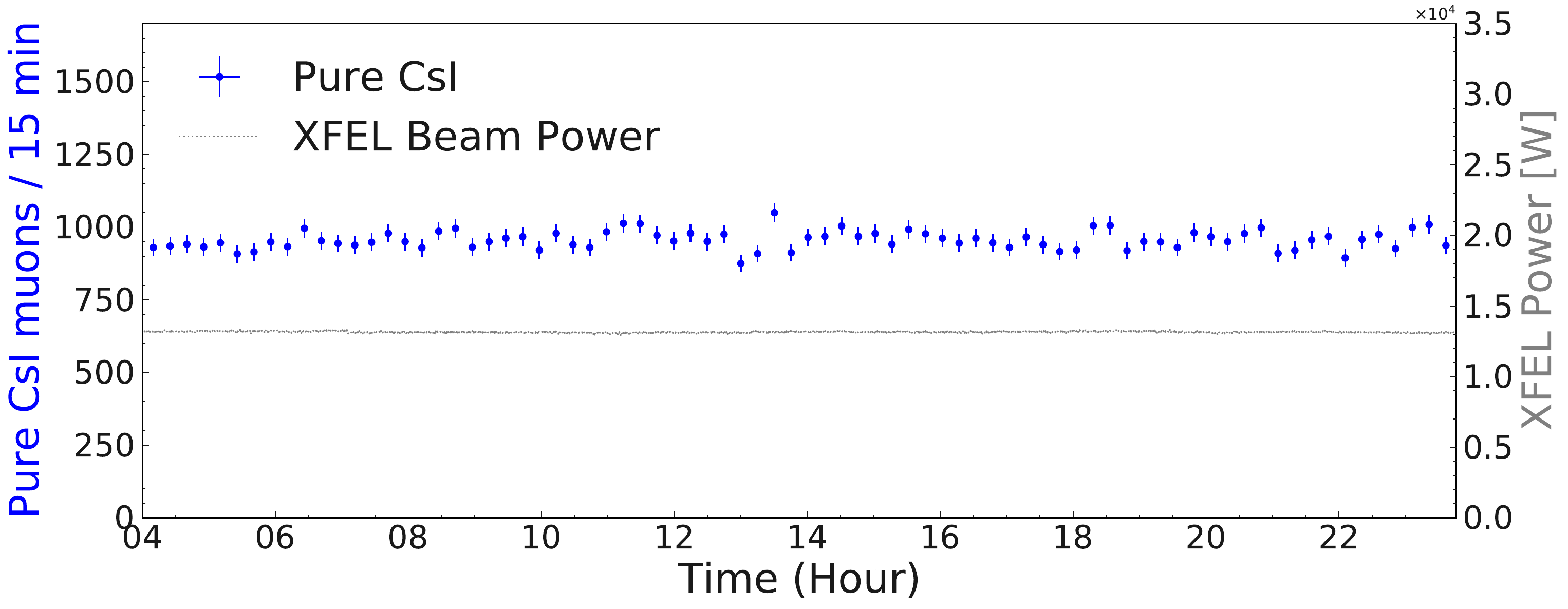}
  \caption{}
\end{subfigure}

\begin{subfigure}{.75\textwidth}
  \centering
  \includegraphics[width=1\linewidth]{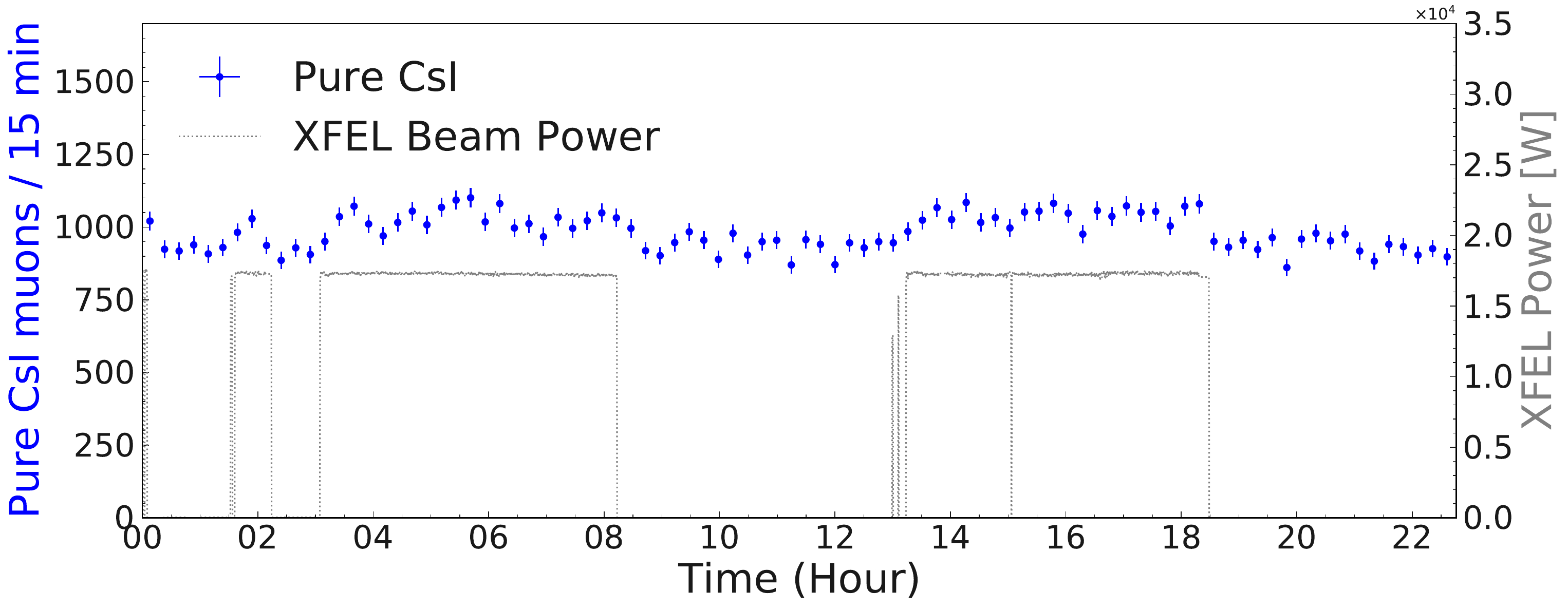}
  \caption{}
\end{subfigure}

\begin{subfigure}{.75\textwidth}
  \centering
  \includegraphics[width=1\linewidth]{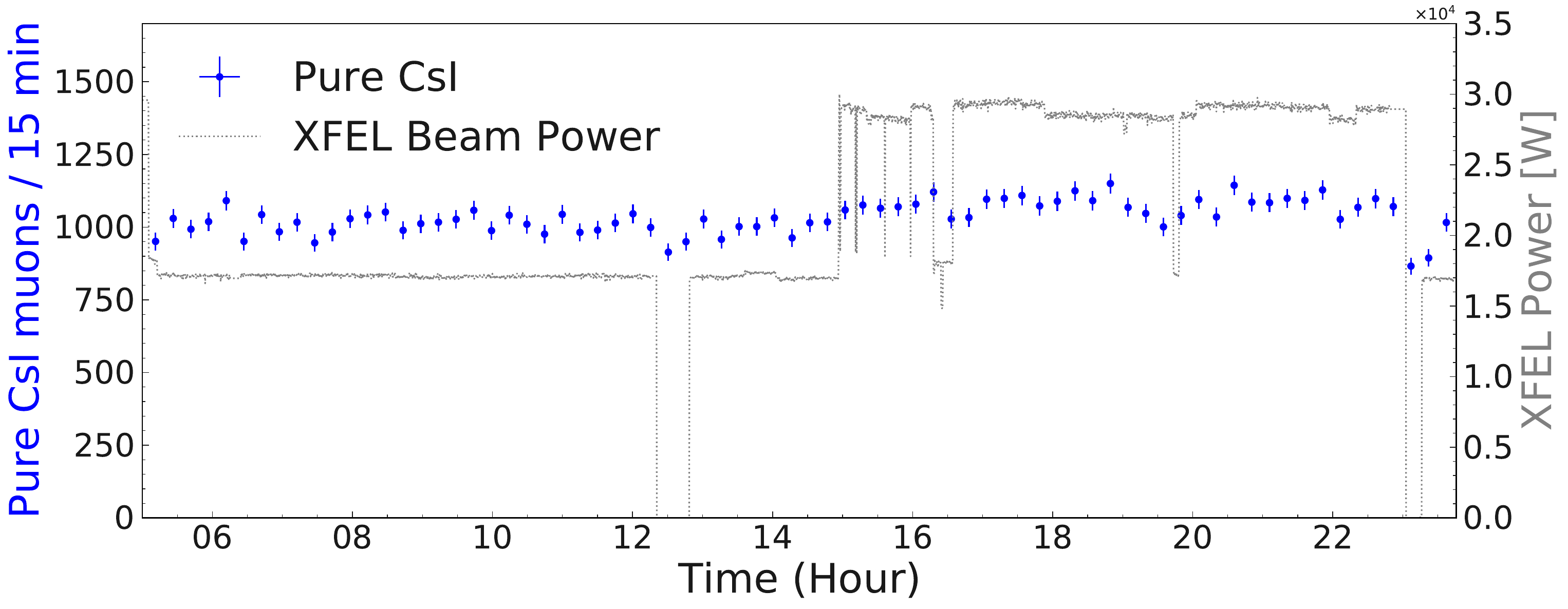}
  \caption{}
\end{subfigure}

\begin{subfigure}{.75\textwidth}
  \centering
  \includegraphics[width=1\linewidth]{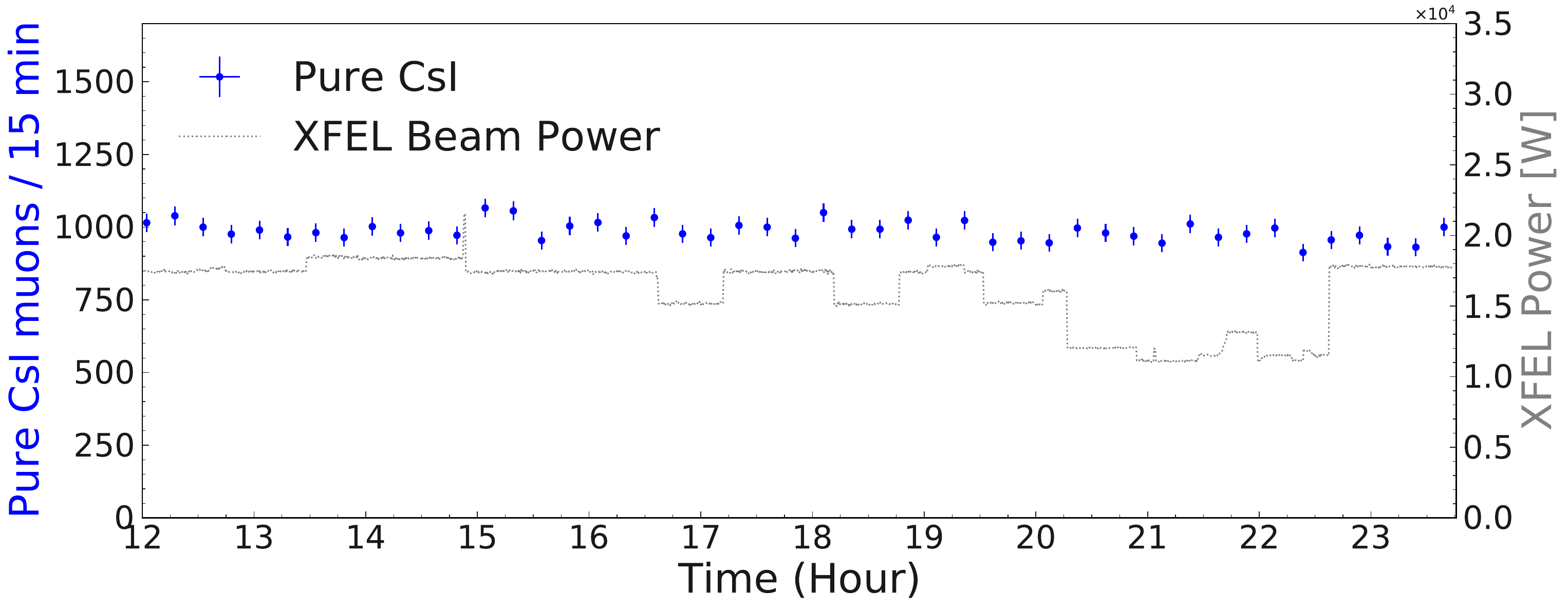}
  \caption{}
\end{subfigure}

\caption{Muon rates measured using pulse shape discrimination with pure CsI for different time periods and compared to the \xfel power.  Muons are expected to arise from cosmic rays and not from \xfel operation.  The weak correlation of the measured muon rate with \xfel power is due the presence of neutron interactions in \purecsi that result in a muon-like pulse shape.}
\label{fig_photonMonitor}
\end{figure}

\noindent applied and can be adjusted depending on the application.  Figures \ref{fig_neutronMonitor_short1} and \ref{fig_neutronMonitor_short2} show two XFEL operation periods when the beam power was changing over a relatively short time scale.  In both cases the pure CsI neutron rate closely follows the beam power trends with less fluctuations than the \pandora system.

The results in Figure \ref{fig_neutronMonitor} demonstrate that using pulse shape discrimination the pure CsI detector can isolate the neutron flux also in presence of a background cosmic muon flux leaving energy deposits with photon-like pulse shapes.  Another potential application of the pure CsI detector system could be to monitor an energetic photon flux while the detector is in a neutron background.  To demonstrate this application a 99\% neutron rejection charge-ratio threshold is defined to classify non-neutron energy deposits in the pure CsI.  An energy deposit above 12 MeV and the 99\% neutron rejection charge-ratio threshold is classified as a muon.  During \xfel operation the absolute rates of muons and neutron interactions in the \purecsi crystals are similar as shown in Figure \ref{fig_EnergyHistograms_difference}.  In Figure \ref{fig_photonMonitor} the values for the non-neutron rates during the same XFEL operational periods as Figure  \ref{fig_neutronMonitor} are shown.  In all cases the recorded non-neutron rate is observed to be approximately stable, independent of the \xfel beam power.  The weak correlation with the \xfel beam power that is observed is from the subset of neutron interactions observed previously in Figure \ref{fig_chargeRatio_1DHistos}, which result in a charge-ratio similar to the muon ionization.  The results in Figure  \ref{fig_photonMonitor} demonstrate that pulse shape discrimination in pure CsI can also effectively identify photon/muon energy deposits in an energetic neutron background.

\section{Conclusion}
\label{sec_conclusion}

The scintillation response of \purecsi to energy deposits from $\sim$100\,MeV neutrons was studied and compared to energy deposits of the same magnitude from cosmic muons. 
Using a charge-ratio method for pulse shape discrimination energy deposits in \purecsi from the energetic neutrons could be distinguished from the cosmic muon energy deposits, which produced photon-like pulse shapes.  The energy deposits from neutron inelastic interactions are observed to produce a lower charge-ratio in \purecsi relative to muons.  This indicates a slower scintillation emission time for the energy deposits from neutron relative to cosmic muons.  The short time window that achieved the optimal pulse shape discrimination in \purecsi was measured to be $100\pm10$\,ns.

The pulse shape discrimination capabilities of \purecsi was qualitatively compared to \csitl. Similar to \csitl, \purecsi was able to separate energy deposits from energetic neutrons vs. muons, however, \purecsi was unable to resolve the specific secondary particles emitted from the neutron inelastic interactions, which is possible with \csitl. In both \csitl and \purecsi approximately 20\% of neutrons produced a waveform with a charge-ratio that was similar to the muon energy deposits. 

The ability for a $5\times5\times30$ cm$^3$ pure CsI crystal to isolate energetic neutron interactions with high purity and moderate efficiency by using pulse shape discrimination was demonstrated. The \purecsi neutron rate was observed to closely follow the \xfel beam power trends.  The energetic neutron monitoring performance was similar to the \pandora commercial radiation monitoring system.  The pulse shape discrimination capabilities of \purecsi in combination with its fast timing and radiation hardness relative to \csitl, makes it an excellent candidate for use in future particle physics calorimeters and radiation detectors.  Beyond the applications in particle and nuclear physics, energetic neutron identification with pure CsI and CsI(Tl) using pulse shape discrimination could additionally be applied in hadron therapy neutron monitoring systems \cite{HOWELL2016249}.  Future studies could also investigate and compare the energetic neutron identification performance of pure CsI and CsI(Tl) with other inorganic scintillators that have demonstrated alpha particle vs. photon/electron identification using pulse shape discrimination, such as CsI(Na) \cite{DINCA2002141}, BaF$_2$ \cite{DINCA2002141}, and PbWO$_4$ \cite{Bardelli:2007di}. 

The results presented used a charge-ratio to perform pulse shape discrimination, however more advanced waveform analysis techniques, such as template fits with optimum filtering \cite{Gatti:1986cw}, and neural networks, could potentially  allow for improved pulse shape discrimination performance.  These techniques could further be implemented on field-programmable gate arrays to perform real-time neutron and photon detection and identification with \purecsi in high radiation environments.  Additionally, crystal-by-crystal variations of the pure CsI and CsI(Tl) fast neutron response should be evaluated to investigate any potential dependence of the pulse shape discrimination capabilities on crystal properties such as dopant and impurity concentration.  The demonstrated pulse shape discrimination capabilities of \purecsi could allow for improvements in calorimeter-based photon vs neutral-hadron as well as lepton vs charged-hadron identification at future particle physics experiments.

\section*{Acknowledgements}
\label{sec:acknowledge}
We would like to thank A.\,Leuschner for providing us with PANDORA data and  K.\,Gadow for help with the mechanical design. We are grateful to C.\,Niebuhr for comments on a draft version of this manuscript. This work is funded by the Deutsche Forschungsgemeinschaft (DFG) under Project  No.\,420484612, under Germany's Excellence Strategy - EXC 2121 ``Quantum Universe'' -  390833306, the Natural Sciences and Engineering Research Council of Canada, and by PIER under Project No.\,PIF-2020-05.
We acknowledge DESY (Hamburg, Germany), a member of the Helmholtz Association HGF, for the provision of experimental facilities. Parts of this research were carried out at \xfel facilities and we thank A.\,Leuschner, S.\,Schreiber, and N.\,Tesch for assistance.

% \appendix

 \bibliographystyle{elsarticle-num} 
 \bibliography{main}

\begin{thebibliography}{10}
\expandafter\ifx\csname url\endcsname\relax
  \def\url#1{\texttt{#1}}\fi
\expandafter\ifx\csname urlprefix\endcsname\relax\def\urlprefix{URL }\fi
\expandafter\ifx\csname href\endcsname\relax
  \def\href#1#2{#2} \def\path#1{#1}\fi

\bibitem{Sugiyama:2021ltp}
Y.~Sugiyama, et~al., {Pulse shape discrimination of photons and neutrons in the
  energy range of 0.1 \textendash{} 2\,GeV with the KOTO un-doped CsI
  calorimeter}, Nucl. Instrum. Meth. A 987 (2021) 164825.

\bibitem{Longo:2020zqt}
S.~Longo, et~al., {CsI(Tl) pulse shape discrimination with the Belle II
  electromagnetic calorimeter as a novel method to improve particle
  identification at electron\textendash{}positron colliders}, Nucl. Instrum.
  Meth. A 982 (2020) 164562.
\newblock \href {http://arxiv.org/abs/2007.09642} {\path{arXiv:2007.09642}}.

\bibitem{Longo:2018uyj}
S.~Longo, J.~M. Roney, {Hadronic vs Electromagnetic Pulse Shape Discrimination
  in CsI(Tl) for High Energy Physics Experiments}, JINST 13~(03) (2018) P03018.
\newblock \href {http://arxiv.org/abs/1801.07774} {\path{arXiv:1801.07774}}.

\bibitem{abe2010belle}
{T. Abe, et al.}, {Belle II Technical Design Report} (2010).
\newblock \href {http://arxiv.org/abs/1011.0352} {\path{arXiv:1011.0352}}.

\bibitem{Belle-II:2018jsg}
W.~Altmannshofer, et~al., {The Belle II Physics Book}, PTEP 2019~(12) (2019)
  123C01, [Erratum: PTEP 2020, 029201 (2020)].
\newblock \href {http://arxiv.org/abs/1808.10567} {\path{arXiv:1808.10567}},
  \href {https://doi.org/10.1093/ptep/ptz106} {\path{doi:10.1093/ptep/ptz106}}.

\bibitem{Barlini_2020}
{S Barlini et al.}, {{FAZIA}: a new performing detector for charged particles},
  Journal of Physics: Conference Series 1561~(1) (2020) 012003.
\newblock \href {https://doi.org/10.1088/1742-6596/1561/1/012003}
  {\path{doi:10.1088/1742-6596/1561/1/012003}}.

\bibitem{KUNDU2019162411}
{S. Kundu et al.}, {ChAKRA : The high resolution charged particle detector
  array at VECC}, Nucl. Instrum. Meth. A 943 (2019) 162411.
\newblock \href {https://doi.org/https://doi.org/10.1016/j.nima.2019.162411}
  {\path{doi:https://doi.org/10.1016/j.nima.2019.162411}}.

\bibitem{Storey1958}
{R. S. Storey, W. Jack and A. Ward}, {The Fluorescent Decay of {CsI}(Tl) for
  Particles of Different Ionization Density}, Proceedings of the Physical
  Society 72~(1) (1958) 1--8.
\newblock \href {https://doi.org/10.1088/0370-1328/72/1/302}
  {\path{doi:10.1088/0370-1328/72/1/302}}.

\bibitem{BARTLE199954}
{C.M. Bartle and R.C. Haight}, Small inorganic scintillators as neutron
  detectors, Nucl. Instrum. Meth. A 422~(1) (1999) 54--58.
\newblock \href {https://doi.org/https://doi.org/10.1016/S0168-9002(98)01062-6}
  {\path{doi:https://doi.org/10.1016/S0168-9002(98)01062-6}}.

\bibitem{DINCA2002141}
{L.E. Dinca et al.}, {Alpha–gamma pulse shape discrimination in CsI:Tl,
  CsI:Na and BaF2 scintillators}, Nucl. Instrum. Meth. A 486~(1) (2002)
  141--145, proceedings of the 6th International Conference on Inorganic
  Scintillators and their Use in Scientific and Industrial Applications.
\newblock \href {https://doi.org/https://doi.org/10.1016/S0168-9002(02)00691-5}
  {\path{doi:https://doi.org/10.1016/S0168-9002(02)00691-5}}.

\bibitem{pureCsIRadHard2016}
{A. Boyarintsev et al.}, {Study of radiation hardness of pure {CsI} crystals
  for Belle-{II} calorimeter} 11~(03) (2016) P03013--P03013.
\newblock \href {https://doi.org/10.1088/1748-0221/11/03/p03013}
  {\path{doi:10.1088/1748-0221/11/03/p03013}}.

\bibitem{KUBOTA1988275}
{S. Kubota et al.}, {A new scintillation material: Pure CsI with 10 ns decay
  time}, Nucl. Instrum. Meth. A 268~(1) (1988) 275--277.
\newblock \href {https://doi.org/https://doi.org/10.1016/0168-9002(88)90619-5}
  {\path{doi:https://doi.org/10.1016/0168-9002(88)90619-5}}.

\bibitem{Aihara_FastCalor}
{H. Aihara et al.}, {Fast Calorimeter on the Pure CsI Crystals for the Modern
  $e^+e^-$ Super Factories}, Physical Society of Japan, 2019, p. 5 pages.
\newblock \href {https://doi.org/10.7566/JPSCP.27.012010}
  {\path{doi:10.7566/JPSCP.27.012010}}.

\bibitem{PIENU_PhysRevLett.115.071801}
{A. Aguilar-Arevalo et al.}, Improved measurement of the
  $\ensuremath{\pi}\ensuremath{\rightarrow}\mathrm{e}\ensuremath{\nu}$
  branching ratio, Phys. Rev. Lett. 115 (2015) 071801.
\newblock \href {https://doi.org/10.1103/PhysRevLett.115.071801}
  {\path{doi:10.1103/PhysRevLett.115.071801}}.

\bibitem{PIBETA_2004}
{E. Frlez et al.}, {Design, commissioning and performance of the PIBETA
  detector at PSI}, Nucl. Instrum. Meth. A 526~(3) (2004) 300–347.
\newblock \href {https://doi.org/10.1016/j.nima.2004.03.137}
  {\path{doi:10.1016/j.nima.2004.03.137}}.

\bibitem{mu2e_2018}
{N. Atanov et al.}, {The Mu2e undoped {CsI} crystal calorimeter}, JINST 13~(02)
  (2018) C02037--C02037.
\newblock \href {https://doi.org/10.1088/1748-0221/13/02/c02037}
  {\path{doi:10.1088/1748-0221/13/02/c02037}}.

\bibitem{Barniakov:2019zhx}
A.~Y. Barniakov, {The Super Charm-Tau Factory in Novosibirsk}, PoS
  LeptonPhoton2019 (2019) 062.
\newblock \href {https://doi.org/10.22323/1.367.0062}
  {\path{doi:10.22323/1.367.0062}}.

\bibitem{EIDELMAN2015238}
{S. Eidelman}, {Project of the Super-tau-charm Factory in Novosibirsk}, Nuclear
  and Particle Physics Proceedings 260 (2015) 238--241, the 13th International
  Workshop on Tau Lepton Physics.
\newblock \href
  {https://doi.org/https://doi.org/10.1016/j.nuclphysbps.2015.02.050}
  {\path{doi:https://doi.org/10.1016/j.nuclphysbps.2015.02.050}}.

\bibitem{PMTdatasheet}
Hamamatsu, {PMT R329-02, Datasheet},
  \url{{https://www.hamamatsu.com/eu/en/product/type/R329-02/index.html}}.

\bibitem{DigitzerDatasheet}
CAEN, {DT5730 / DT5730S, Datasheet}, \url{
  https://www.caen.it/products/dt5730/}.

\bibitem{KLETT20101242}
{A. Klett, A. Leuschner and N. Tesch}, A dose meter for pulsed neutron fields,
  Radiation Measurements 45~(10) (2010) 1242--1244, {PROCEEDINGS OF THE 11TH
  SYMPOSIUM ON NEUTRON AND ION DOSIMETRY}.
\newblock \href {https://doi.org/https://doi.org/10.1016/j.radmeas.2010.06.008}
  {\path{doi:https://doi.org/10.1016/j.radmeas.2010.06.008}}.

\bibitem{PandoraDatasheet}
{Berthold Technologies}, {LB 6419, Datasheet},
  \url{{https://www.berthold.com/en/radiation-protection/products/dose-and-dose-rate/neutron-and-gamma-dose-rate-monitor-lb-6419/}}.

\bibitem{PANDORApaper}
{A. Leuschner}, {Dose rate measurements around the electron extraction at
  FLASH}\url{https://radsynch15.desy.de/sites/site_radsync15/content/e268576/e268578/infoboxContent268622/Session03_radsynch15_Leuschner.pdf}
  (2015).

\bibitem{DESYInternalReport}
{T. T. Liang et al.}, \href{https://bib-pubdb1.desy.de/record/476764}{{{D}ose
  rates calculations from electron beam losses on {XS}1 dump at {XFEL}}}, Tech.
  Rep. DESY-D3–130 (2021).
\newblock \href {https://doi.org/10.3204/PUBDB-2022-01873}
  {\path{doi:10.3204/PUBDB-2022-01873}}.
\newline\urlprefix\url{https://bib-pubdb1.desy.de/record/476764}

\bibitem{scipy}
E.~Jones, T.~Oliphant, P.~Peterson, et~al.,
  \href{http://www.scipy.org/}{{SciPy}: Open source scientific tools for
  {Python}} (2001--).
\newline\urlprefix\url{http://www.scipy.org/}

\bibitem{2020SciPy-NMeth}
{P. Virtanen et al}, {{{SciPy} 1.0: Fundamental Algorithms for Scientific
  Computing in Python}}, {Nature Methods} 17 (2020) 261--272.
\newblock \href {https://doi.org/10.1038/s41592-019-0686-2}
  {\path{doi:10.1038/s41592-019-0686-2}}.

\bibitem{caleb_w_fink_2021_5104856}
{C. W. Fink and S. L. Watkins},
  \href{https://doi.org/10.5281/zenodo.5104856}{{QETpy}} (July 2021).
\newblock \href {https://doi.org/10.5281/zenodo.5104856}
  {\path{doi:10.5281/zenodo.5104856}}.
\newline\urlprefix\url{https://doi.org/10.5281/zenodo.5104856}

\bibitem{Zyla:2020zbs}
P.~Zyla, et~al., {Review of Particle Physics}, PTEP 2020~(8) (2020) 083C01.
\newblock \href {https://doi.org/10.1093/ptep/ptaa104}
  {\path{doi:10.1093/ptep/ptaa104}}.

\bibitem{MuniraThesis}
{M. Khan}, \href{https://docs.belle2.org/record/2977}{{Laboratory and
  Simulation studies of Pulse Shape Discrimination in pure CsI and CsI(Tl)}},
  Master's thesis, Universit{\"a}t Hamburg (2021).
\newline\urlprefix\url{https://docs.belle2.org/record/2977}

\bibitem{knoll2010radiation}
G.~Knoll, Radiation Detection and Measurement, Wiley, 2010.

\bibitem{MCLEAN2006793}
{T. D. McLean et al.}, {CHELSI: Recent developments in the design and
  performance of a high-energy neutron spectrometer}, Nucl. Instrum. Meth. A
  562~(2) (2006) 793--796, proceedings of the 7th International Conference on
  Accelerator Applications.
\newblock \href {https://doi.org/https://doi.org/10.1016/j.nima.2006.02.057}
  {\path{doi:https://doi.org/10.1016/j.nima.2006.02.057}}.

\bibitem{HOWELL2016249}
{R. M. Howell et al.}, {Measured Neutron Spectra and Dose Equivalents From a
  Mevion Single-Room, Passively Scattered Proton System Used for Craniospinal
  Irradiation}, International Journal of Radiation Oncology*Biology*Physics
  95~(1) (2016) 249--257, {Particle Therapy Special Edition}.
\newblock \href {https://doi.org/https://doi.org/10.1016/j.ijrobp.2015.12.356}
  {\path{doi:https://doi.org/10.1016/j.ijrobp.2015.12.356}}.

\bibitem{Bardelli:2007di}
L.~Bardelli, et~al., {Pulse-shape discrimination with PbWO(4) crystal
  scintillators}, Nucl. Instrum. Meth. A 584 (2008) 129--134.
\newblock \href {http://arxiv.org/abs/0706.2422} {\path{arXiv:0706.2422}},
  \href {https://doi.org/10.1016/j.nima.2007.10.021}
  {\path{doi:10.1016/j.nima.2007.10.021}}.

\bibitem{Gatti:1986cw}
E.~Gatti, P.~F. Manfredi, {Processing the Signals From Solid State Detectors in
  Elementary Particle Physics}, Riv. Nuovo Cim. 9N1 (1986) 1--146.
\newblock \href {https://doi.org/10.1007/BF02822156}
  {\path{doi:10.1007/BF02822156}}.

\end{thebibliography}

\appendix
\section{Appendix: Distribution of waveform energies before and after pre-selection}

\begin{figure}[H]

\begin{subfigure}{.49\textwidth}
  \centering
\includegraphics[width=1\textwidth]{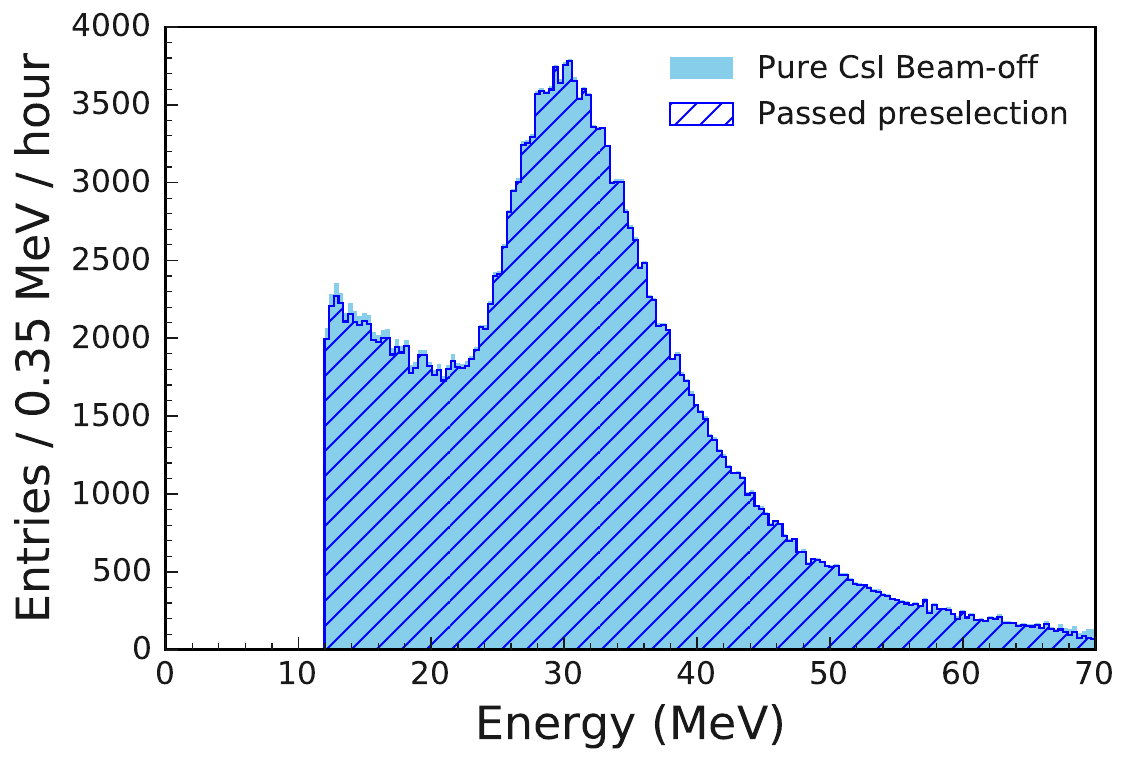}\\
  \caption{}
\end{subfigure}
\begin{subfigure}{.49\textwidth}
  \centering
\includegraphics[width=1\textwidth]{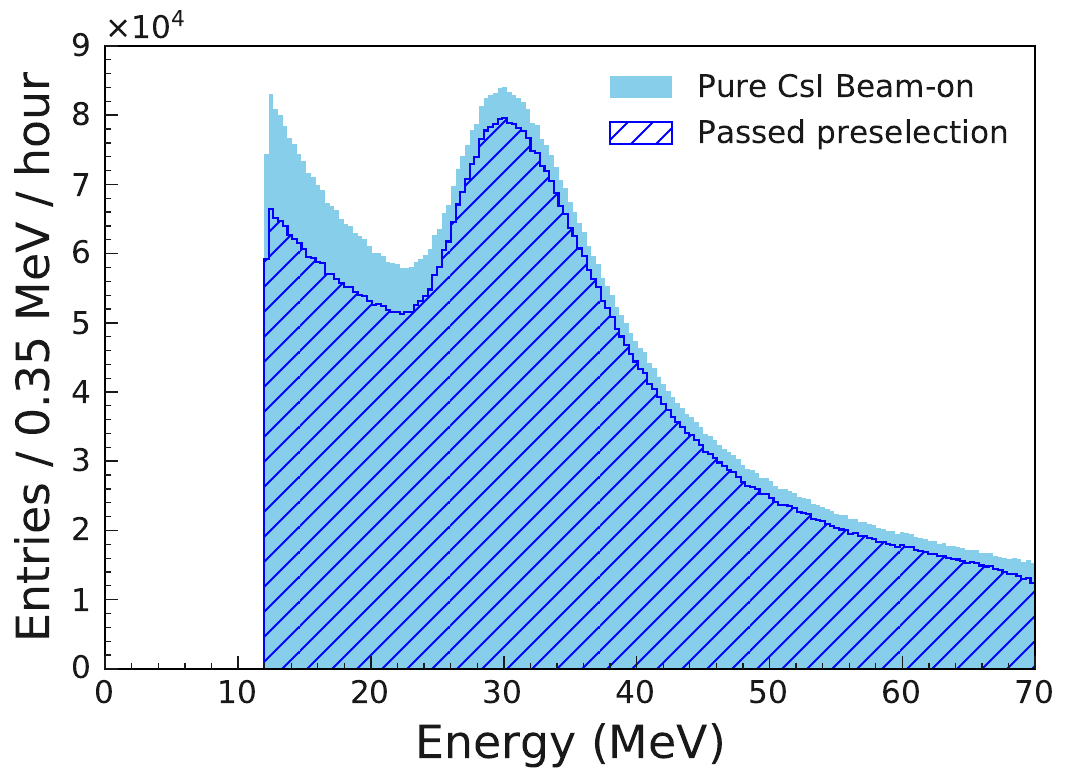}\\
  \caption{}
\end{subfigure}

\centering
\begin{subfigure}{.49\textwidth}
  \includegraphics[width=1\linewidth]{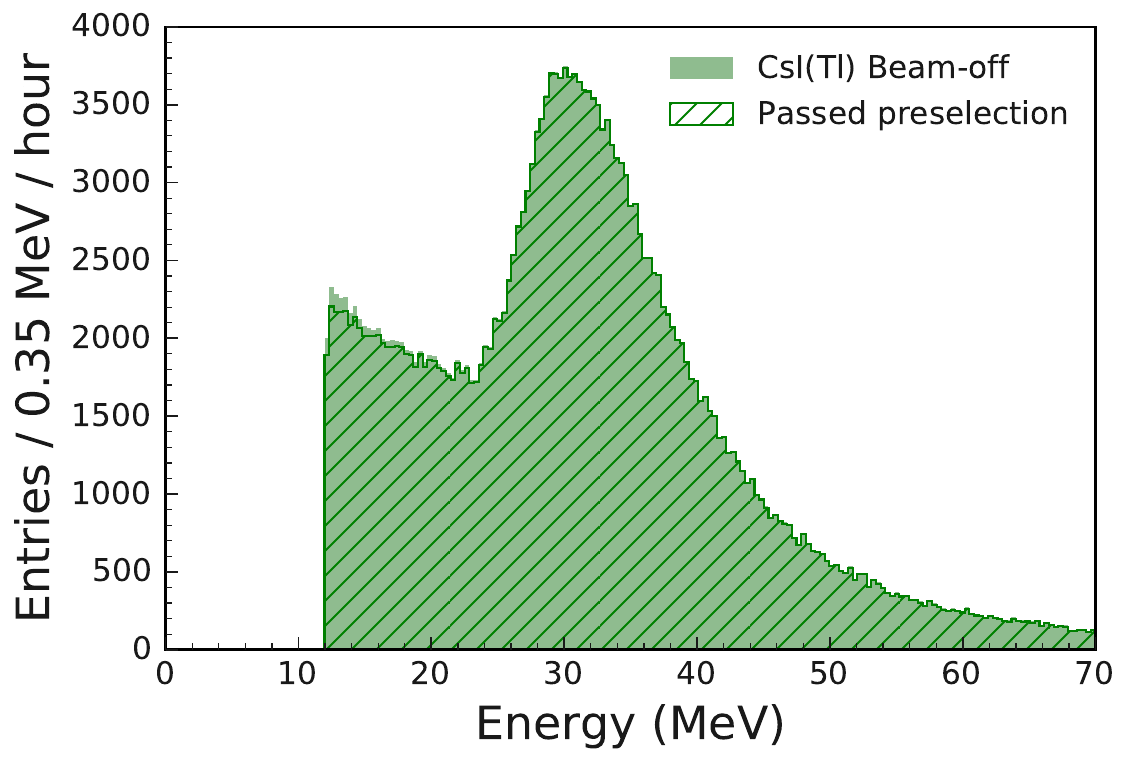}
  \caption{}
\end{subfigure}
\begin{subfigure}{.49\textwidth}
  \includegraphics[width=1\linewidth]{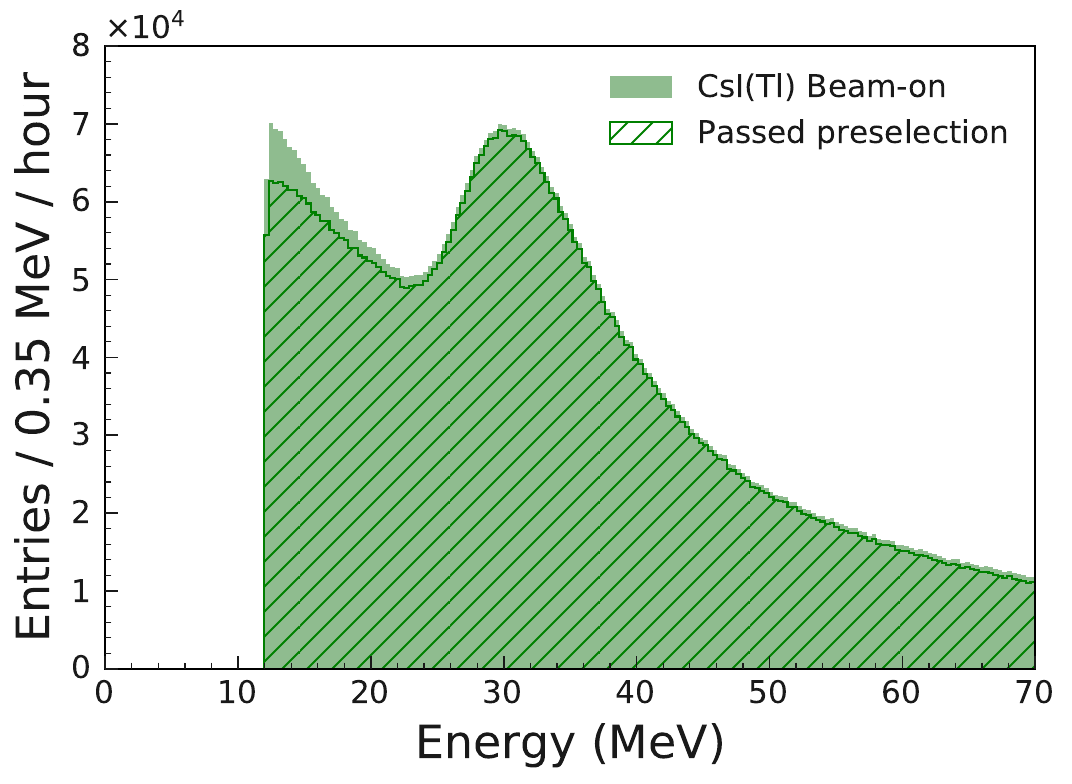}
  \caption{}
\end{subfigure}

\caption{Distribution of the total energy of waveforms recorded in the a) Pure CsI beam-off,  b) Pure CsI beam-on, c) CsI(Tl) beam-off and d) CsI(Tl) beam-on datasets before and after the pre-selection is applied.}
\label{fig_EnergyHistograms_preselection}
\end{figure}

The waveform pre-selection detailed in Section \ref{sec2_ExpSetup} is applied to remove waveforms exceeding the maximum ADC value and waveforms with multiple pile-up in the 30\,$\upmu$s time window.  Figure \ref{fig_EnergyHistograms_preselection} shows the energy distributions of the waveforms before and after the pre-selection is applied.  Due to the different pile-up criteria described in Section \ref{sec2_ExpSetup}, the pre-selection is observed to remove more waveforms in the pure CsI beam-on compared to the CsI(Tl).

\end{document}